\documentclass[10pt,
  prl,
  twocolumn,
  aps,
  amssymb,
  amsmath,
  longbibliography,
  showpacs,
  superscriptaddress,
  floatfix,
  colorlinks,
  linkcolor=blue,
  citecolor=blue,
  anchorcolor=blue
]{revtex4-2}

\usepackage{comment}
\usepackage{xcolor}
\usepackage{graphicx}
\usepackage{dcolumn}
\usepackage{bm}
\usepackage{hyperref}
\usepackage{mathtools}
\usepackage[mathlines]{lineno} 
\usepackage{units}
\usepackage{balance}

\newcommand{\figOne}{
    \begin{figure*}[t]
    \includegraphics[width=\textwidth]{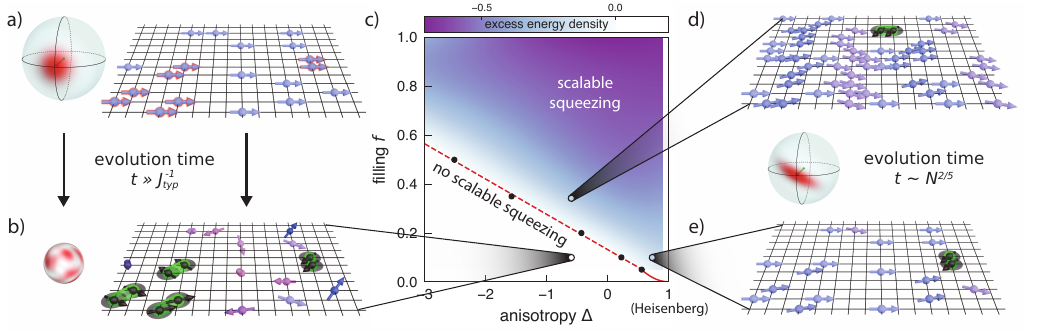}
    \caption{\emph{Scalable squeezing with disordered dipoles in 2D.}
    (a) Depicts an example of a disordered spin configuration in two dimensions. 
    For strong disorder (here $f=0.1$), near-neighbor “dimers” (red glow) are strongly excited when prepared in an $x$-polarized product state. 
    For such a product state, the Wigner distribution of the collective spin on the Bloch sphere (red) is concentrated around the $+x$ direction with $O(\sqrt{N})$ projection noise.
    (b) Under evolution with $H_\mathrm{XXZ}$, dimers relax toward their local ground states (black arrows), releasing energy that can heat the system above the XY ordering temperature $T_\mathrm{c}$.
    (c) Time evolution with $H_\mathrm{XXZ}$ yields scalable spin squeezing when the energy density of the initial $x$-state ($E_x$) is below the energy density of the critical state ($E_\mathrm{c}$), i.e.~when the excess energy density $(E_x-E_\mathrm{c})/|E_x| < 0$ (density plot).
    The phase boundary is obtained from QMC (black points, red dashed guide for the eye) and for $f \to 0$, from analytic arguments (red solid).
    As $f \to 0$, the region of scalable spin squeezing is localized to a regime near the Heisenberg point ($\Delta=1$). 
    From the non-squeezing phase, scalable squeezing can be restored by (d) increasing the filling toward the uniform case $f=1$, which reduces the hierarchy of energy scales between strong dimer bonds and typical interactions; or (e) increasing $\Delta$ toward the Heisenberg point, which lowers the dimer excitation energy. Optimal squeezing is believed to occur at timescales $t \sim N^{2/5}$~\cite{block_scalable_2024}.
    }\label{fig:squeezing}
    \end{figure*}
}

\newcommand{\figTwo}{
    \begin{figure}[t]
    \includegraphics[width=\columnwidth]{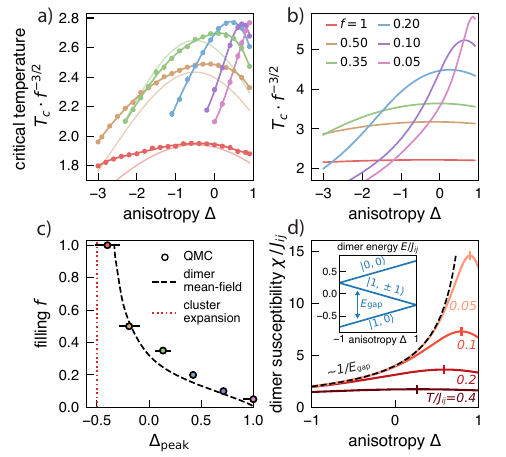}
    \caption{\emph{Thermodynamic phase diagram of XY order.} 
    (a) Critical temperature $T_\mathrm{c}$ from QMC (solid lines) versus Ising anisotropy $\Delta$ for several filling fractions $f$ (legend shared with panel (b)).
    Temperature is rescaled by $f^{-3/2}$ to account for the change in the typical nearest-neighbor dipolar interaction strength as one dilutes the system.
    Note that $T_\mathrm{c}$ extends further below the numerically accessible points shown.
    For the three largest fillings, the first-order mean-field cluster expansion is also shown (faint lines). 
    (b) The analogous $T_\mathrm{c}$ curves for the dimer mean-field ansatz [Eq.~\eqref{eq:MF_dimers}], which captures full intra-dimer correlations. 
    (c) Depicts the anisotropy $\Delta_{\rm peak}$ which maximizes $T_\mathrm{c}$ as extracted from QMC for different filling fractions $f$ (colored circles), and from the mean-field theory with dimer corrections (dashed black). In contrast, the cluster expansion yields $\Delta_{\rm peak} = -0.5$ for all $f$ (dotted red).
    (d) The critical temperature $T_\mathrm{c}$ is larger when the dimers are more susceptible to XY ordering; this is quantified for a given dimer (with internal coupling $J_{ij}$) by its magnetic susceptibility $\chi$ to a transverse field from the other spins [Eq.~\eqref{eq:chi}], shown here at various anisotropies $\Delta$ and temperatures $T$. At $T\!\approx\!E_{\mathrm{gap}}$ the susceptibility crosses over from Curie behavior $\chi\sim 1/T$ to $\chi\sim 1/E_{\mathrm{gap}}$. 
    (inset) Shows the energy spectrum of a dimer as a function of the anisotropy. 
    }\label{fig:Tc}
    \end{figure}
}

\newcommand{\figThree}{
    \begin{figure}[t]
    \includegraphics[width=\columnwidth]{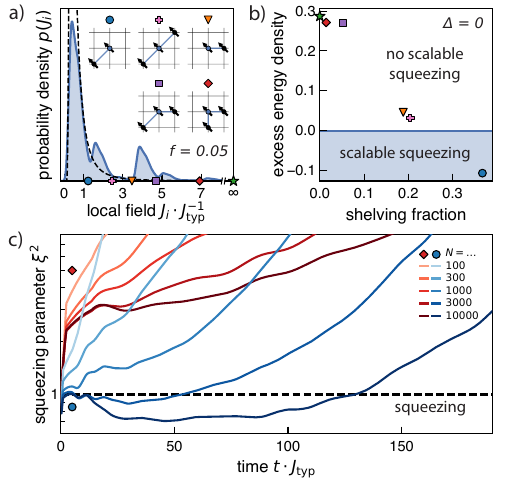}
    \caption{
    \emph{Scalable squeezing via the shelving of dimers.} 
    (a) Rescaled distribution $p(J_i)$ of local fields $J_i := \sum_{j\ne i} J_{ij}$ for $f=0.05$ (solid blue) and $f \to 0$ (dashed black)~\cite{SM}. 
    The markers on the $x$ axis label peaks (to their right) in the $f=0.05$ distribution, each corresponding to the fields indicated by the local motifs in the insets (top right).
    By shelving all spins with $J_i$ to the right of a given marker, (b) one lowers the excess energy density of the initial $x$-state $(E_x-E_\mathrm{c})/|E_x|$ by an amount roughly proportional to the fraction of spins shelved, as demonstrated by QMC for $\Delta = 0$.
    (c) Dynamical simulations using cDTWA confirm the squeezing phase diagram derived from QMC in (b).
    Increasing the fraction of spins shelved from 1.5\% (red) to 35\% (blue) leads to an optimal squeezing parameter $\xi^2$ below $1$ and decreasing with post-shelving system size $N$.
    We have filtered out fast oscillations in $\xi^2(t)$ due to dimers~\cite{SM}.
    }\label{fig:shelving}
    \end{figure}
}

\begin{document}

\title{Theory of Scalable Spin Squeezing with Disordered Quantum Dipoles}

\author{Avi Kaplan-Lipkin}\thanks{These authors contributed equally to this work.}
	\affiliation{Department of Physics, Harvard University, Cambridge, Massachusetts 02138, USA}

	\author{Philip J. D. Crowley}\thanks{These authors contributed equally to this work.}
	\affiliation{Department of Physics, Harvard University, Cambridge, Massachusetts 02138, USA}

    \affiliation{Department of Physics and Astronomy, Michigan State University, East Lansing, Michigan 48824, USA}

    \author{Jonathan N. Hall\'en}

    	\affiliation{Department of Physics, Harvard University, Cambridge, Massachusetts 02138, USA}

    \affiliation{Department of Physics, Boston University, Boston, Massachusetts 02215, USA}

     \author{Zilin Wang}

    	\affiliation{Department of Physics, Harvard University, Cambridge, Massachusetts 02138, USA}

         \author{Weijie Wu}

    	\affiliation{Department of Physics, Harvard University, Cambridge, Massachusetts 02138, USA}

         \author{Sabrina Chern}

    	\affiliation{Department of Physics, Harvard University, Cambridge, Massachusetts 02138, USA}

           \author{Chris R. Laumann}

    	\affiliation{Department of Physics, Harvard University, Cambridge, Massachusetts 02138, USA}

    \affiliation{Department of Physics, Boston University, Boston, Massachusetts 02215, USA}

        \author{Lode Pollet}

    	\affiliation{Department of Physics and Arnold Sommerfeld Center for Theoretical Physics (ASC), Ludwig-Maximilians-Universität München, Theresienstrasse 37, D-80333 München, Germany}

    \affiliation{Munich Center for Quantum Science and Technology (MCQST), Schellingstrasse 4, D-80799 München, Germany}

     \author{Norman Y. Yao}

     	\affiliation{Department of Physics, Harvard University, Cambridge, Massachusetts 02138, USA}

\begin{abstract} 

Spin squeezed entanglement enables metrological precision beyond the classical limit. 
Understood through the lens of continuous symmetry breaking, dipolar spin systems exhibit the remarkable ability to generate spin squeezing via their intrinsic quench dynamics. 
To date, this understanding has primarily focused on lattice spin systems; in practice however, dipolar spin systems---ranging from ultracold molecules to nuclear spin ensembles and solid-state color centers---often exhibit significant amounts of positional disorder.
Here, we develop a theory for scalable spin squeezing in a two-dimensional randomly diluted lattice of quantum dipoles, which naturally realize a dipolar XXZ model. 
Via extensive quantum Monte Carlo simulations, we map out the phase diagram for finite-temperature XY order, and by extension scalable spin squeezing, as a function of both disorder and Ising anisotropy.
As the disorder increases, we find that scalable spin squeezing survives only near the Heisenberg point. 
We show that this behavior is due to the presence of rare tightly-coupled dimers, which effectively heat the system post-quench. 
In the case of strongly-interacting nitrogen-vacancy centers in diamond, we demonstrate that an experimentally feasible strategy to decouple the problematic dimers from the dynamics is sufficient to enable scalable spin squeezing.
\end{abstract}

\maketitle

The probabilistic nature of quantum mechanical measurements leads to a fundamental noise floor on the precision of any sensor~\cite{giovannetti_advances_2011, degen_quantum_2017, pezze_quantum_2018}.
The classic strategy for improving this precision is to use an ensemble of $N$ independent sensors, yielding a suppression of the quantum projection noise that scales as $\sim1/\sqrt{N}$.
Remarkably, quantum entanglement in the form of \emph{spin squeezing} can be used to reshape the quantum projection noise and achieve sensitivities $\sim 1/ N^\gamma$ surpassing the classical limit~\cite{wineland_spin_1992, kitagawa_squeezed_1993, meyer_experimental_2001, ma_quantum_2011}.
In theory, \emph{scalable} spin squeezing ($\gamma > \nicefrac{1}{2}$) enables measurement sensitivities reaching all the way to the Heisenberg limit $\sim 1/N$.
Spin squeezing with all-to-all interactions has been studied extensively~\cite{esteve_squeezing_2008, leroux_implementation_2010,gross_nonlinear_2010,riedel_atomchipbased_2010,hamley_spinnematic_2012,ockeloen_quantum_2013,bohnet_quantum_2016a,hosten_measurement_2016,braverman_nearunitary_2019} with experimental demonstrations of up to $\approx 20$ dB of metrological gain~\cite{hosten_measurement_2016}.
However, recent theoretical insights have demonstrated that scalable squeezing can be generated in a significantly broader class of \emph{locally}-interacting models~\cite{perlin_spin_2020,comparin_robust_2022,comparin_scalable_2022,block_scalable_2024,koyluoglu_squeezing_2025a}.
These findings establish finite-temperature continuous-symmetry-breaking as a key ingredient for scalable squeezing, developing a remarkable connection between the system's dynamical and equilibrium properties~\cite{comparin_robust_2022,comparin_scalable_2022,block_scalable_2024}. 
This theoretical progress has had an almost immediate impact on experiments---in large part because it suggests that the experimentally ubiquitous setting of two-dimensional quantum dipoles can support scalable spin squeezing~\cite{valtolina_dipolar_2020,comparin_multipartite_2022,chen_continuous_2023a,su_dipolar_2023,gong_coherent_2023,hughes_strongly_2025,franke_quantumenhanced_2023}.
Indeed, squeezing dynamics have recently been observed with electric dipoles in Rydberg atom arrays~\cite{bornet_scalable_2023} and ultracold polar molecules~\cite{bilitewski_dynamical_2021}, with magnetic dipoles in a quantum gas microscope~\cite{douglas_spin_2025}, and with ensembles of solid-state color centers~\cite{wu_spin_2025}.

The solid-state setting stands out from the other synthetic platforms in which spin squeezing has been demonstrated.
In stark contrast to the latter, solid-state color centers are routinely used for a wide variety of practical sensing applications, ranging from nanoscale magnetometry~\cite{maze_nanoscale_2008a,schirhagl_nitrogenvacancy_2014,kolkowitz_probing_2015,casola_probing_2018,jenkins_imaging_2022,rovny_nanoscale_2024,he_probing_2025} to \emph{in vivo} thermometry~\cite{kucsko_nanometrescale_2013,lesage_optical_2013,aslam_quantum_2023}.
However, an intrinsic challenge associated with color centers is the presence of strong disorder owing to their random positions within the solid.
This motivates the following essential question: 
What is the fate of scalable spin squeezing in the presence of strong positional disorder?

\figOne

In this Letter, we present and analyze a theory for scalable spin squeezing in two-dimensional ensembles of disordered quantum dipoles.
In the presence of a strong, out-of-plane, quantizing field, such dipoles are described by the paradigmatic spin-$\nicefrac{1}{2}$ XXZ model~\cite{zu_emergent_2021}:
\begin{equation}\label{eq:H_XXZ}
    H_\mathrm{XXZ}=-\sum_{i<j} J_{ij}\left(s_i^xs_j^x+s_i^ys_j^y+\Delta s_i^zs_j^z\right),
\end{equation}
where $J_{ij} = J/r_{ij}^3$, $\vec{s}_i$ are spin operators, and $\Delta$ characterizes the Ising anisotropy (below, $J=1$). 
Positional disorder is parameterized via the filling fraction $f$ of independently and randomly occupied sites on an underlying lattice [Fig.~\ref{fig:squeezing}]~\cite{alonso2010monte,andresen2014existence,kwasigroch_synchronization_2017,zhang_quantum_2018,gannarelli2012contribution}; $f = 1$ corresponds to the disorder-free case (a uniform lattice), while $f \to 0$ corresponds to strong disorder (a highly diluted lattice)~\footnote{We note that limited cuts of the phase diagram have been explored using QMC for $\Delta = 0, -3.5$~\cite{zhang_quantum_2018} and via semi-classical methods at low filling for $\Delta=0$~\cite{kwasigroch_synchronization_2017}.}.

Our main results are threefold.
First, we consider the conventional protocol for generating spin squeezing via quench dynamics from an initial, spin-polarized product state. 
Scalable squeezing requires the energy density of the initial product state to be below the critical energy density for ferromagnetic XY order in $H_\mathrm{XXZ}$~\cite{frerot_entanglement_2017,sbierski_magnetism_2024,block_scalable_2024}, or by Legendre transform, the effective temperature to be below the critical temperature.
Accordingly, we determine, via extensive path-integral quantum Monte Carlo (QMC), the critical temperature for XY order as a function of both the filling fraction $f$ and the Ising anisotropy $\Delta$. 
This allows us to map out the phase diagram for scalable spin squeezing and establish the critical disorder strength above which scalable squeezing does not occur [Fig.~\hyperref[fig:squeezing]{\ref*{fig:squeezing}c}]. 

Second, we explore a surprising and highly consequential feature of the phase diagram [Fig.~\hyperref[fig:squeezing]{\ref*{fig:squeezing}c}]: As the disorder increases ($f \to 0$), scalable spin squeezing survives in a vanishingly narrow range of $\Delta$ near the Heisenberg point ($\Delta=1$). 
We show that this behavior is due to the presence of rare ``dimers'' (i.e.~pairs of spins with atypically large coupling), which emerge with strong disorder. 
Two effects are at play: (i) away from the Heisenberg point, it is energetically favorable for the dimers to occupy the unpolarized triplet state, rather than participate in the XY ordering, and (ii) when initialized in a spin-polarized state, the dimers are extremely ``hot'' and release energy to the surrounding spins during the quench dynamics.
We attempt to analytically capture these effects via both a perturbative cluster expansion and a non-perturbative treatment of dimer correlations, successfully reproducing the key features of our quantum Monte Carlo simulations. 
Finally, returning to the experimental setting of solid-state color centers, we analyze the prospects of scalable spin squeezing in strongly-interacting ensembles of nitrogen-vacancy (NV) defects in diamond.
We find that, for current generation samples~\cite{hughes_strongly_2025}, strong positional disorder prohibits the generation of scalable spin squeezing via a simple quench from a polarized initial state. 
Recent experiments have proposed and implemented strategies for effectively ``shelving'' dimers and eliminating their impact on the squeezing dynamics~\cite{wu_spin_2025}.
Motivated by this, we perform a new set of modified QMC simulations with varying degrees of dimer shelving and demonstrate that such shelving enables the recovery of scalable spin squeezing.

\emph{Critical temperature for XY order}.---%
Consider dilute dipoles governed by $H_\textrm{XXZ}$ on a square lattice.
In order to determine the regime of scalable spin squeezing, we compute the critical temperature, $T_\textrm{c}$, for XY order as a function of $\Delta$, at a variety of filling fractions down to $f = 0.05$  [Fig.~\hyperref[fig:Tc]{\ref*{fig:Tc}a}].
We use disorder-averaged quantum Monte Carlo with worm-type updates on an imaginary time path integral representation \cite{sadoune_efficient_2022,SM}.
As one dilutes the system, the typical distance between dipoles increases, $r_{\rm typ} \sim 1/\sqrt{f}$, leading to an interaction strength that scales as $J_{\rm typ} = 1/r_{\rm typ}^3 \sim f^{3/2}$.
It is therefore natural to rescale all energies---including the critical temperature---by $f^{-3/2}$.

In the disorder-free case [$f = 1$, red curve Fig.~\hyperref[fig:Tc]{\ref*{fig:Tc}a}], $T_\textrm{c}$ exhibits relatively weak $\Delta$-dependence with a slight peak at $\Delta_{\rm peak} \approx -0.5$. 
The introduction of disorder leads to a marked change in this behavior: the critical temperature becomes sharply peaked as a function of $\Delta$ and the location of this peak drifts strongly toward the Heisenberg point, $\Delta = 1$ [Fig.~\hyperref[fig:Tc]{\ref*{fig:Tc}a}].
This implies that positional disorder drastically suppresses finite temperature XY order, which at moderate temperatures confines it to a narrowing region around $\Delta = 1$.

To turn our map of critical temperatures into a phase diagram for scalable spin squeezing, let us recall the quench protocol.
We prepare an initial, $x$-polarized product state $|\psi_x\rangle =  \left | \rightarrow \cdots \rightarrow \right \rangle_x$ and then evolve under $H_\textrm{XXZ}$ up to the time at which squeezing is optimized.
If $|\psi_x\rangle$ thermalizes to a temperature $T_x < T_\textrm{c}$ (or equivalently its energy density satisfies $E_x < E_\textrm{c}$), the XY order is stable upon dynamical evolution and the system exhibits scalable squeezing~\cite{block_scalable_2024}.
This immediately yields the phase diagram depicted in Fig.~\hyperref[fig:squeezing]{\ref*{fig:squeezing}c}. 

Notably, the region of scalable squeezing inherits the behavior of XY order---it is sharply compressed toward $\Delta =1$ as the filling fraction decreases.
In other words, scalable squeezing is in principle possible for arbitrarily strong positional disorder provided $\Delta$ is sufficiently close to the Heisenberg point.
However, there is an issue:  the optimal timescale  diverges as $\sim (1 - \Delta)^{-1}$ in this limit.
To see this, note that precisely at the Heisenberg point, $|\psi_x\rangle$ is a ground state exhibiting no dynamics; away from this point, the squeezing dynamics are generated by the SU(2)-symmetry-breaking terms in the Hamiltonian, with an energy scale $\sim(1-\Delta)$. 
Taken together, these constraints lead us to the conclusion that scalable squeezing with dilute dipoles is infeasible using the conventional quench protocol. 
This motivates the following question: What is the underlying physics that causes the scalable squeezing region in Fig.~\hyperref[fig:squeezing]{\ref*{fig:squeezing}c} to vanish at low $f$, and can one develop strategies to resurrect it?

\figTwo

\emph{Mean-field cluster expansion}.---%
In order to gain intuition, we consider a standard mean-field picture, which turns out to work reasonably near $f=1$ but fails in the limit of strong disorder.
Our starting point is a self-consistent variational mean-field theory, with $H_{\rm MF} = -h \sum_i s_i^x$~\cite{yeomans_statistical_1992,SM}. 
The mean-field condition on the critical temperature for XY order is,
\begin{equation}\label{eq:mf_Tc_condition}
     \sum_{ij}\chi_{ij} = \sum_{ijkl} \chi_{ij}\, J_{jk}\, \chi_{kl},
\end{equation}
with $\chi_{ij}$ the mean-field transverse susceptibility of spin $i$ to a field applied on spin $j$ (see Supplemental Material (SM)~\cite{SM}).
Under $H_{\rm MF}$, the susceptibility takes the simple diagonal form, $\chi_{ij} = \delta_{ij}/4T$.
The mean-field critical temperature is then given by $T_\textrm{c}^{\rm MF} = (4N)^{-1}\sum_{ij}J_{ij}= 2.25\ldots \times f$. 
This estimate fails to capture nearly all of the qualitative properties of the phase diagram in Fig.~\hyperref[fig:Tc]{\ref*{fig:Tc}a}: It exhibits no $\Delta$ dependence and scales as $T_\textrm{c}^{\rm MF} \sim f$, rather than $T_\textrm{c} \sim f^{3/2}$.
A natural extension is to consider a high-temperature cluster expansion~\cite{kubo_generalized_1962,Oitmaa_Hamer_Zheng_2006}, which enables one to capture few-body quantum correlations near $T_\mathrm{c}$. 
Working with the modified variational ansatz, $H_\lambda = (1-\lambda) H_{\rm MF} +\lambda H_{\rm XXZ}$, we compute $T_\mathrm{c}$ to leading order in $\lambda$~\cite{SM} and obtain the solid curves in Fig.~\hyperref[fig:Tc]{\ref*{fig:Tc}a}.
This approach qualitatively reproduces the $\Delta$-dependence in $T_\textrm{c}$ for weak disorder. 
But there is much to be desired.
For example, the peak in $T_\textrm{c}$ stays pinned at $\Delta_{\rm peak} = -0.5$ independent of $f$, rather than shifting toward the Heisenberg point. 
It is clear that the physics at strong disorder is simply not being captured by this analysis.
Indeed, by treating all spins equally in the ansatz, the analysis neglects the impact of clusters of strongly-coupled spins, which is particularly important at strong dilution. 
These randomly occurring clusters consist of spins whose pairwise separation $r_{ij}$ is atypically small, $r_{ij} \ll r_{\rm typ} \sim 1/\sqrt{f}$. 
In the limit of strong disorder ($f\to 0$), the density of such clusters containing $n$ spins scales as $f^n$, naturally motivating us to focus our attention on the effect of dimers~[Fig.~\hyperref[fig:squeezing]{\ref*{fig:squeezing}a}]. 
\emph{Understanding the impact of dimers}.---%
The strong disorder regime requires taking into account the physics of strongly-coupled dimers.
The dimers have an unmagnetized local ground state $|S=1,m=0\rangle$, with an energy gap
\begin{equation}\label{eq:Egap}
    E_{\rm gap} = J_{ij}(1-\Delta)/2
\end{equation}
to their other triplet states (see inset, Fig.~\hyperref[fig:Tc]{\ref*{fig:Tc}d}).
At temperatures of order $J_{\rm typ}$ ($\ll J_{ij}$), the dimers are ``frozen out'' ($E_{\rm gap} \gg T$) and do not participate in the magnetic ordering.
However, $E_{\rm gap}$ vanishes as the anisotropy $\Delta$ is tuned towards the Heisenberg point.
This is reflected in the magnetic susceptibility of a dimer as a large peak that develops at temperatures $T \ll J_{ij}$ (see Fig.~\hyperref[fig:Tc]{\ref*{fig:Tc}d} and the End Matter).
Thus, near the Heisenberg point, the dimers do in fact participate in XY order, increasing the density of unfrozen spins and raising the overall $T_\mathrm{c}$.

To quantitatively account for the role of dimers, we extend our previous mean-field theory to include their quantum correlations: 
\begin{equation}\label{eq:MF_dimers}
    H_\mathrm{MF}^{\rm (d)}=-h\sum_i s_i^x - \sum_{\mathclap{(ij) \in \mathrm{dimers}}}J_{ij}\left(s_i^xs_j^x+s_i^ys_j^y+\Delta s_i^zs_j^z\right),
\end{equation}
where $(ij)\in{\rm dimers}$ indicates that spins $i$ and $j$ form a dimer.
Numerically, we identify strongly-coupled dimers using a real-space matching algorithm~\cite{braemer_cluster_2024, SM}.
Under this new mean-field Hamiltonian, the susceptibility $\chi_{ij}$ in Eq.~\eqref{eq:mf_Tc_condition} has off-diagonal contributions from the dimer correlations (see End Matter and SM~\cite{SM}).
The resulting critical temperature as a function of both $f$ and $\Delta$ are depicted in Fig.~\hyperref[fig:Tc]{\ref*{fig:Tc}b} and capture the qualitative features observed in the QMC [Fig.~\hyperref[fig:Tc]{\ref*{fig:Tc}a}].
As illustrated in Fig.~\hyperref[fig:Tc]{\ref*{fig:Tc}c}, we find that $\Delta_{\rm peak}$ extracted from the dimer mean-field theory exhibits excellent agreement with that extracted from the QMC. 

\figThree

The structure of the scalable squeezing phase diagram [Fig.~\hyperref[fig:squeezing]{\ref*{fig:squeezing}c}] can be understood from the dimer picture as well.
We begin at the Heisenberg point $\Delta = 1$, where the $x$-state is a many-body ground state, and thus $E_x < E_\mathrm{c}$ for all $f$.
Away from this point, a quench from the $x$-state leaves highly excited dimers, which relax to their ground states and release energy $\delta E$ into the surrounding spins.
For a \emph{single} dimer (see Eq.~\eqref{eq:MF_dimers}) direct calculation yields
\begin{equation}
    \delta E := E_x - E_\mathrm{gs} =-|h| + \sqrt{h^2 + \tfrac14 E_{\rm gap}^2}
\end{equation}
where $h \sim J_\mathrm{typ} \sim f^{3/2}$ is the local field due to other spins, and $E_{\rm gap}$ is given by Eq.~\eqref{eq:Egap}.
We first discuss the dilute limit of $f\to0$.
In this limit, the dimer interactions satisfy $E_{\rm gap} \gg h$, and hence $\delta E \sim E_{\rm gap}\sim J_{ij}(1-\Delta)$.
The total energy release is dominated by the most strongly coupled dimers ($J_{ij}=1$), even though their fraction vanishes as $\sim f$.
Their relaxation thus collectively raises the energy of the remaining spins by $\sim (1-\Delta) \times f$ per spin.
When this energy exceeds the interaction scale between typical spins $h\sim f^{3/2}$, which stabilizes XY order, it heats the system above its ordering temperature ($T_x > T_\mathrm{c}$) and prevents squeezing.
Equating these two scales yields the asymptotic phase boundary
$1-\Delta \sim \sqrt{f}$ as $f \to 0$.
This form is consistent with the shrinking squeezing region observed in the QMC [Fig.~\hyperref[fig:squeezing]{\ref*{fig:squeezing}c}], but it fails to capture the observed linear phase boundary.
To understand the linear phase boundary, we consider the pre-asymptotic, intermediate-$f$ regime, where $J_{\rm typ} = O(1)$.
We then have $h \gtrsim E_{\rm gap}/2$, and perturbative analysis yields $\delta E\propto E_{\rm gap}^2 / h$.
Inserting this into the same heating criterion yields a crossover boundary $1-\Delta \sim f$, in agreement with the QMC data in Fig.~\hyperref[fig:squeezing]{\ref*{fig:squeezing}c}.
For further details see the SM~\cite{SM}.

\emph{Scalable spin squeezing in solid-state color center ensembles}.---%
Our analysis applies to a variety of experimental platforms featuring disordered dipoles, including, for example, ultracold polar molecules in optical lattices ($f \sim 0.2$)~\cite{moses_creation_2015} and ensembles of solid-state color centers ($f \sim 0.001$)~\cite{gong_coherent_2023,hughes_strongly_2025,he_probing_2025}.
Here, we focus on the latter, and investigate the prospects for generating \emph{scalable} spin squeezing in strongly-interacting ensembles of nitrogen-vacancy (NV) color centers in diamond~\cite{doherty_nitrogenvacancy_2013}; NV centers represent a leading platform for quantum sensing~\cite{maze_nanoscale_2008a,schirhagl_nitrogenvacancy_2014,kolkowitz_probing_2015,casola_probing_2018,jenkins_imaging_2022,rovny_nanoscale_2024,gao_signal_2025} and spin squeezing has recently been demonstrated~\cite{wu_spin_2025}.
However, owing to the natural anisotropy ($\Delta = -1$)~\footnote{We note that the anisotropy can be modified via Floquet engineering to any $\Delta > -1$~\cite{choi_robust_2020, zhou_quantum_2020}, as experimentally demonstrated in~\cite{martin_controlling_2023}.} and  strong disorder, such squeezing was far from the scalable regime.

Having identified dimers as the key barrier to scalable squeezing, we investigate an experimentally-motivated strategy for overcoming this challenge via 
dynamically ``shelving'' a fraction of the NV centers~\cite{wu_spin_2025}.
The effect of this shelving is to remove strongly-coupled dimers from the squeezing dynamics. 
In practice, one can identify the dimer spins via the anomalously large local fields they experience in a uniformly polarized system, given by $J_i := \sum_{j\ne i} J_{ij}$.
Accordingly, we generate an ensemble of disorder configurations, and for each, compute the set of $J_i$ and remove all spins with $J_i>J_0$.
For non-vanishing $f$, the distribution over $J_i$ has several features arising from the lattice constant, which naturally suggest places to set our cutoff, $J_0$.
Each feature and its corresponding cutoff are indicated by the geometric markers in Fig.~\hyperref[fig:shelving]{\ref*{fig:shelving}a}.
These features vanish in the $f\to0$ limit [black dashed line, Fig.~\hyperref[fig:shelving]{\ref*{fig:shelving}a}].

To demonstrate the feasibility of shelving for recovering scalable spin squeezing, we perform QMC ($\Delta = 0.0$, $f = 0.05$) for these cutoffs and compute the critical energy density, $E_\textrm{c}$.
By comparing these critical energy densities to those of the $x$-polarized state, $E_x$, we find that approximately 25\% of the spins must be shelved in order to recover scalable squeezing [Fig.~\hyperref[fig:shelving]{\ref*{fig:shelving}b}].
A recent experiment on a dense NV ensemble \cite{wu_spin_2025} demonstrated this approximate amount of shelving.

While Fig.~\hyperref[fig:shelving]{\ref*{fig:shelving}b} suggests that scalable squeezing can be achieved with sufficient shelving, it remains to directly verify this from the quench dynamics of the $x$-polarized state.
To this end, we directly simulate the quench dynamics for different shelving cutoffs using cluster DTWA (Discrete Truncated Wigner Approximation)~\cite{schachenmayer_manybody_2015a,braemer_cluster_2024,SM}.
Squeezing is quantified by the Wineland squeezing parameter, $\xi^2\equiv N {\rm{min}}_{\hat{n}\perp\hat{x}}{\rm{Var}}\left[\hat{n}\cdot\vec{S}\right]/\langle S^x \rangle^2$, with $\vec{S}=\sum_i \vec{s}_i$ the total spin operator and $\xi^2<1$ indicating metrological squeezing~\cite{wineland_squeezed_1994}.
For minimal shelving ($1.5\%$ of spins, red curves in Fig.~\hyperref[fig:shelving]{\ref*{fig:shelving}c}), $\xi^2$ increases rapidly at short times and does not exhibit spin squeezing even for large system sizes, $N\sim 10^4$.
By contrast, for moderate shelving where $E_x < E_\mathrm{c}$ ($35\%$ of spins, blue curves in Fig.~\hyperref[fig:shelving]{\ref*{fig:shelving}c}), we find that the dynamics of $\xi^2$ exhibit a significant change as a function of system size, and scalable squeezing is indeed recovered for $N \gtrsim 10^3$. 

\emph{Conclusion and Outlook}.---%
Our work opens the door to a number of intriguing future directions.
First, while we have focused on dipolar interactions, a number of experimental quantum simulation platforms exhibit shorter range interactions in 2D (e.g.~van der Waals~\cite{eckner_realizing_2023,hines_spin_2023a} or contact interactions~\cite{lee_observation_2025}) where only algebraic long-range order is present at finite temperatures~\cite{cuccoli_twodimensional_1995,giachetti_berezinskiikosterlitzthouless_2022}. 
It would be interesting to investigate the role of disorder on spin squeezing in these settings.
Second, beyond quenched disorder, it is also natural to study dynamical disorder, which is intrinsic to itinerant dipolar systems~\cite{bilitewski_dynamical_2021,wellnitz_spin_2024a,carroll_observation_2025,douglas_spin_2025}.
Finally, our experimental proposal for overcoming disorder has involved engineering the Hamiltonian in order to raise the critical temperature for XY order.
However, one can also explore an alternative route based upon adiabatically preparing an initial (slightly-entangled) state with a lower energy density.

\emph{Acknowledgments}.---%
We gratefully acknowledge the insights of and discussions with Maxwell Block, Emily Davis, Che Liu, Markus M\"uller, Rahul Nandkishore, Tasuku Ono, Haopu Yang, and Bingtian Ye.
This work is supported in part by the U.S.\ Department of Energy BES Grant No. DE-SC0019241 (mean-field cluster expansion calculations) and the Army Research Office through the MURI program Grant No.~W911NF-20-1-0136 (quantum Monte Carlo simulations).
L.P.\ acknowledges support from  the Deutsche Forschungsgemeinschaft (DFG, German Research Foundation) under Germany's Excellence Strategy -- EXC-2111 -- project number 390814868. 
C.R.L.\ acknowledges the Gutzwiller Fellowship at Max Planck Institute for the Physics of Complex Systems and the partial support of SFB 1143 and ct.qmat.
J.N.H.\ acknowledges support from The Sweden-America Foundation.
N.Y.Y.\ acknowledges support from a Brown Investigator award.

\nocite{wallerberger2018updatedcorelibrariesalps,GAENKO2017235,prokofev_exact_1998,allen_computer_2017,defenu_longrange_2023,lemieux_control_2017}

\bibliography{refs}

\section{End Matter}

\emph{Magnetic susceptibility of a dimer}.---%
At strong disorder, we have attributed the $\Delta$-dependence of the critical temperature $T_\mathrm{c}$ to the role of strongly-coupled dimers. In fact, $T_\mathrm{c}$ inherits its peak from that of the dimer susceptibility, as seen in Fig.~\hyperref[fig:Tc]{\ref*{fig:Tc}d}. This dimer susceptibility enters at a mean-field level via Eq.~\eqref{eq:mf_Tc_condition}, and in the following we compute it explicitly and study its limiting behaviors.

Consider the Hamiltonian for the dimer interaction under a field,
\begin{equation}\label{eq:Hij}
    H_{ij} = - J_{ij} (s_i^x s_j^x + s_i^y s_j^y + \Delta s_i^z s_j^z) + h \left(s_i^x + s_j^x\right).
\end{equation}
We are interested in the dimer's magnetic susceptibility $\chi$, which represents its susceptibility to XY ordering with the field $h$ from the rest of the system at the critical temperature. This is given by,
\begin{equation}\label{eq:chi}
    \chi = \left.\frac{\langle s_i^x + s_j^x\rangle}{\partial h}\right|_{h=0} = \frac{2\left(1-e^{-E_{\rm gap}/T}\right)/E_{\rm gap}}{1+e^{-J_{ij}/T}+2e^{-E_{\rm gap}/T}}.
\end{equation}
with the expectation value taken in the thermal distribution according to $H_{ij}$ at temperature $T$. There is implicit $\Delta$ dependence in the energy gap $E_{\rm gap} = (1-\Delta)J_{ij}/2$. Note that in the language of Eq.~\eqref{eq:mf_Tc_condition}, this dimer susceptibility is $\chi = \chi_{ii} + \chi_{ij} + \chi_{ji} + \chi_{jj}$.

For the temperature regime of interest (near $T_\mathrm{c}$), dimers are defined by $J_{ij} \gg J_{\rm typ} \sim T$. In this regime, the highest energy state---the singlet state---has a population $O\left(e^{-J_{ij}/T}\right)$ and may be neglected. The dimers are further split into two classes depending on their energy gap $E_{\rm gap}$:
\begin{enumerate}
    \item Frozen dimers, satisfying $E_{\rm gap} \gg T$. This is the Van Vleck regime ($\chi \sim 1/E_{\rm gap}$) where \emph{virtual} excitations dominate: \[ \chi_{\rm Van\ Vleck} = \frac{2}{E_{\rm gap}}+O\left(e^{-E_{\rm gap}/T}\right)+O\left(e^{-J_{ij}/T}\right).\]
    \item Unfrozen dimers, satisfying $E_{\rm gap} \ll T$. This is the Curie regime ($\chi \sim 1/T$) where \emph{thermal} excitations dominate: \[\chi_{\rm Curie} = \frac{1}{T}\left(\frac{2}{3}+O\left(\frac{E_{\rm gap}}{T}\right)\right)+O\left(e^{-J_{ij}/T}\right).\]
\end{enumerate}

These two regimes are apparent in Fig.~\hyperref[fig:Tc]{\ref*{fig:Tc}d}. A frozen dimer is gapped out from XY ordering, since that would require weight in the higher-energy, thermally-suppressed triplet states. An unfrozen dimer instead behaves as an $S=1$ spin and is perfectly capable of XY ordering. However, if $E_{\rm gap}$ is too small, this polarization is unstable to thermal fluctuations. The competition between freezing when $E_{\rm gap}$ is too large (far below the Heisenberg point) versus thermal mixing when it is too small (very near the Heisenberg point) leads to a peak in $\chi$ that occurs at $E_{\rm gap}/T = 1.0356\dots$. At this point, the the susceptibility attains its maximal value $\chi/\chi_H = 1.0926\dots$, where $\chi_H$ is the value of $\chi$ at the Heisenberg point ($\Delta = 0$).

\end{document}



\title{Supplemental Material}

\maketitle

\tableofcontents

\section{Quantum Monte Carlo}

\subsection{Algorithmic details}

To study the equilibrium properties of $H_{\rm XXZ}$ (Eq. 1 of the main text) at temperature $T$, we employ path-integral quantum Monte Carlo (QMC) with worm type updates, adapted from the codebase in Ref.~\cite{sadoune_efficient_2022}. The code makes use of the ALPSCore libraries \cite{wallerberger2018updatedcorelibrariesalps,GAENKO2017235}, and the QMC algorithm is based on Ref.~\cite{prokofev_exact_1998} with two notable changes.

The first change we have made to the code in Ref.~\cite{sadoune_efficient_2022} is that we use an Ewald summation to calculate the $1/r^3$ interactions of the 2D system~\cite{allen_computer_2017}. The interactions are given by,
\begin{equation}\label{eq:Ewald}
    J_{ij} = \sum_{\vec{n}\in\mathbf{Z}^2} \frac{J}{|\vec{r}_{ij}+L\vec{n}|^3},
\end{equation}
where $\vec{n}\in\mathbf{Z}^2$ runs over all ordered integer pairs $(m,n)$, so that spin $i$ interacts with every periodic image of spin $j$ in the infinitely replicated system. We find that this choice improves convergence compared to open or periodic boundary conditions.

The second change is to average over disorder realizations of the diluted lattice. Given a side length $L$ of an underlying square lattice, and a filling fraction $f$ of independently occupied sites on that underlying lattice, we randomly choose $N=fL^2$ (rounded to the nearest odd integer) lattice sites to populate with spins. We then average over enough disorder realizations to see a converged histogram for the observables of interest. The number of disorder realizations required increases as the filling is decreased, in order to maintain the same uncertainty on the sample mean. For example, for $f=0.5$ each point is averaged over 80 disorder realizations, whereas for $f=0.05$ each point is averaged over 300 disorder realizations.

We note that the algorithm is optimized for the lattice case and is not suitable for the strong disorder limit in which spins are placed randomly in continuous space. In this limit, the worm can easily become ``stuck,'' hopping back and forth between two strongly interacting spins, leading to longer computation times per Monte Carlo step. The lowest filling fraction we were able to reach in this study is therefore $f=0.05$.

\subsection{Extraction of critical temperature}

\begin{figure}[h]
    \centering
    \includegraphics{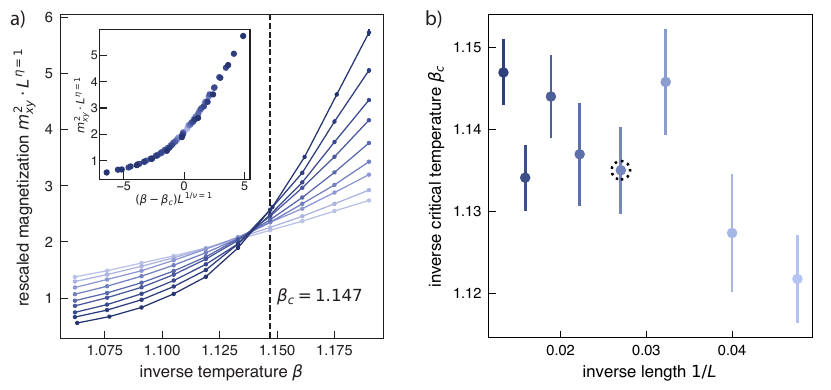}
    \caption{QMC measurement of critical temperature. (a) Extraction of the inverse critical temperature for $f=0.5$, $\Delta = 0.0$. Each data point is averaged over 80 disorder realizations. Darker shades of blue correspond to larger system sizes, across values $L = 21, 25, 31, 37, 45, 53, 63, 75, 89$. (inset) Scaling collapse of the curves according to Eq.~\eqref{eq:scaling}. (b) The drift of $\beta_\mathrm{c}$ from crossings of successive pairs of system sizes in (a). As described in the text, we report $\beta_\mathrm{c}$ of the circled point.}
    \label{fig:magnetization}
\end{figure}

Our task is to find the critical temperature $T_\mathrm{c}$ and corresponding energy density $E_\mathrm{c}$. To do so, we use QMC to compute the in-plane magnetization per spin,
\begin{equation}\label{eq:mperp}
    m_{xy}^2=\frac{1}{N}\sum_{ij}\langle s_i^x s_j^x + s_i^y s_j^y \rangle.
\end{equation}
This magnetization obeys the finite size scaling form,
\begin{equation}\label{eq:scaling}
    m_{xy}^2 L^\eta\sim F\left(L/\xi\right),
\end{equation}
with $F$ a non-universal scaling function and $\xi\sim |\beta-\beta_\mathrm{c}|^{-\nu}$ the correlation length at inverse temperature $\beta = 1/T$. We use the mean-field estimates $\eta=\nu=1$ \cite{defenu_longrange_2023} which do appear to collapse the different system sizes [inset of Fig.~\hyperref[fig:magnetization]{\ref*{fig:magnetization}a}]. At the critical temperature, the correlation length diverges and $m_{xy}^2 L^\eta$ becomes independent of $L$, which gives us a procedure for identifying the critical point from this crossing, as depicted in Fig.~\hyperref[fig:magnetization]{\ref*{fig:magnetization}a}.

\begin{figure}[h]
    \centering
    \includegraphics{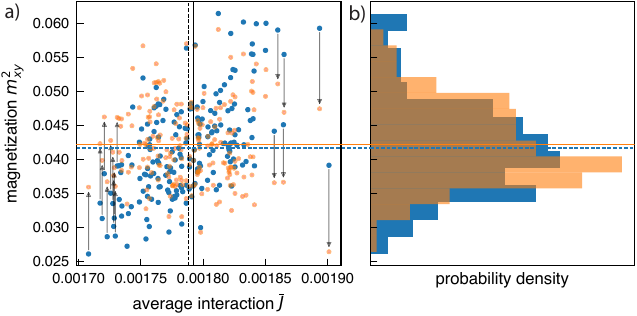}
    \caption{Method of control variates. (a) QMC at $f=0.2$, $\Delta=-0.5$, $L=71$, $\beta=4.5$, showing the relation between $m_{xy}^2$ and the average pairwise interaction $\overline{J}$ from $200$ disorder realizations. Blue points depict the raw data, with blue and black dashed lines showing the sample mean along each axis. The solid black line shows the true mean of $\overline{J}$. This is used to correct to the orange points according to Eq.~\eqref{eq:control_variates}, with new sample mean given by the orange dashed line. Arrows depict the movement of the points with the strongest deviation. (b) Histograms over the data in panel (a) with a shared $y$-axis. The mean shifts by $1\%$, and we use this shifted (orange) mean $\widetilde{m}_{xy}^2$ as our estimate.}
    \label{fig:control_variates}
\end{figure}

We use the method of control variates~\cite{lemieux_control_2017} to improve our estimate of $m_{xy}^2$. Specifically, we observe that the measured magnetization is strongly correlated with the average pairwise interaction $\overline{J}=\frac{1}{N(N-1)}\sum_{j>i}J_{ij}$. While the magnetization must be computed statistically from Monte Carlo disorder realizations, the average of $\langle J_{ij} \rangle$ over disorder realizations is a much lighter computation that can therefore be performed more precisely. If our sample mean (over disorder realizations) of $\overline{J}$ deviates from the true mean, we can use this deviation to correct the sample mean of $m_{xy}^2$ according to its covariance with $\overline{J}$, by using the following modified estimator:
\begin{equation}\label{eq:control_variates}
    \widetilde{m}_{xy}^2= m_{xy}^2 -\frac{\mathrm{Cov}\left(m_{xy}^2,\overline{J}\right)}{\mathrm{Var}\left(\overline{J}\right)}\left(\overline{J}-\left\langle \overline{J} \right\rangle\right),
\end{equation}
where angle bracket averages and (co)variances are taken over disorder averages. In short, this approach improves the convergence of histograms over disorder realizations by correcting for unusually weak or strong interactions (Fig.~\ref{fig:control_variates}).

Let $\beta_\mathrm{c}(N_1, N_2)$ be the crossing point of $m_{xy}^2 L^\eta$ for the two system sizes $N_1<N_2$. As both are increased, we expect the crossing point to converge to the true critical point $\beta_\mathrm{c}$. In practice, $\beta_\mathrm{c}(N_1, N_2)$ drifts with system size, and we consistently observe a non-monotonicity for $N \lesssim 500$. For larger $N$ than this, we find that $\beta_\mathrm{c}$ is converged within error bars. We therefore always report $\beta_\mathrm{c}(N_1=675, N_2 = 1000)$ for the critical point (circled point in Fig.~\hyperref[fig:magnetization]{\ref*{fig:magnetization}b}) and double the error bars---calculated from bootstrapping our disorder samples---to account for systematics.

\begin{figure}[hbt]
    \centering
    \includegraphics{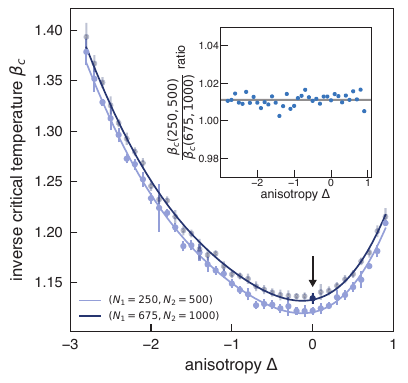}
    \caption{Extrapolation of $\beta_\mathrm{c}$ to larger systems. Shows the inverse critical temperature computed from two smaller system sizes (light blue points) and two larger system sizes (dark blue points). Note that the ratio of $\beta_\mathrm{c}$ for the two curves is independent of $\Delta$ (see inset). The light blue curve is a Gaussian fit to the light blue data, whereas the dark blue curve is the same curve, scaled by the ratio of the two points at $\Delta=0$ (black arrow). The latter curve nevertheless closely follows the true values for all $\Delta$, enabling precise initial guesses for QMC at these system sizes.}
    \label{fig:delta_independence}
\end{figure}

For the large system sizes shown in Fig.~\ref{fig:magnetization}, it is only computationally feasible to search large swaths of $\beta$ for the crossing point at exactly $\Delta=0$. This is because there is then no diagonal energy to be computed for each Monte Carlo move, leading to a significant speedup in runtime. At $\Delta \ne 0$, we rely on an initial guess derived from smaller system sizes to narrow our search. This strategy is shown in Fig.~\ref{fig:delta_independence} and consists of three steps:
\begin{enumerate}
    \item For a range of anisotropies $-3 < \Delta < 1$, we measure $\beta_\mathrm{c}(N_1 = 250, N_2 = 500;\Delta)$, using linear interpolation in $\beta$ to find the crossing point.
    \item At just $\Delta = 0$, we additionally measure $\beta_\mathrm{c}(N_1 = 675, N_2 = 1000;\Delta=0)$. Its ratio to the smaller $N$s provides an initial guess for $\Delta\ne0$ at larger $N$s: we guess $\beta_\mathrm{c}(N_1 = 675, N_2 = 1000;\Delta)$ is near $\dfrac{\beta_\mathrm{c}(N_1 = 675,\, N_2 = 1000;\Delta=0)}{\beta_\mathrm{c}(N_1 = 250,\, N_2 = 500;\Delta=0)}\beta_\mathrm{c}(N_1 = 250, N_2 = 500;\Delta)$
    \item We use these initial guesses to calculate $\beta_\mathrm{c}(N_1 = 675, N_2 = 1000;\Delta)$ as in step 1, but with a narrowed search in $\beta$. We emphasize that only first principle data is used at this point.
\end{enumerate}

\subsection{Extraction of critical energy}

\begin{figure}[hbt]
    \centering
    \includegraphics{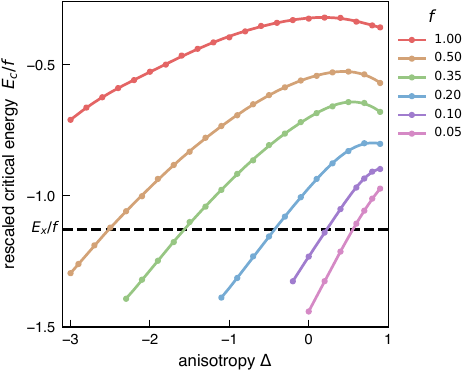}
    \caption{QMC measurement of critical energy. Same data used for Fig.~1c in the main text, where the phase boundary for scalable squeezing comes from the intersection of the QMC critical energy with the $x$-state's energy (black dashed line). The curves shown for $E_\mathrm{c}$ (and similarly $T_\mathrm{c}$ in other plots) are fits with Gaussian processes shown as guides for the eye.}
    \label{fig:energies}
\end{figure}

To get Fig.~1c of the main text, we use linear interpolation from the QMC energy per spin at nearby $\beta$ to calculate $E_\mathrm{c}$ at the measured $\beta_\mathrm{c}$. These critical energies are shown in Fig.~\ref{fig:energies}. At lower temperatures, the Monte Carlo steps take longer, because of the growing window of imaginary time for the worm to explore (of size $\beta$). For each $f$, we therefore only go to low enough temperatures to see $E_\mathrm{c}$ cross below $E_x$.

\section{\label{sec:Ji}Distribution of local fields}

In this section we analyze the distribution of local fields $J_i := \sum_{j\ne i} J_{ij}$ for the diluted lattice model with $J_{ij} = Jr_{ij}^{-3}$. At $f = 0.05$ [Fig.~3a of the main text], the distribution shows several peaks, each associated with a distinct motif of nearby spins on the lattice. These motifs are defined by specific patterns of horizontal, vertical, and diagonal neighbors, which set a minimal \emph{threshold} local field, since couplings to more distant spins only increase $J_i$. The markers on the horizontal axis in Fig.~3a of the main text indicate these threshold fields, whose values are listed in Table~\ref{tab:Ji_motifs}; the random environment beyond these spins is responsible for the shift of the corresponding peaks to larger $J_i$.

\begin{table}[h]
	\centering
    \begin{tblr}{
        colspec={c c c},
        cells={c},   
        hline{2} = {2pt},
        row{1} = {font=\bfseries},
        hlines,
        vlines
    }
        Nearest spins & Threshold $J_i$ & Picture \\
        One diagonal neighbor & $2^{-3/2}=0.353\ldots$ & \motif{2} \\
        Two diagonal neighbors & $2\times 2^{-3/2}=0.707\ldots$ & \motif{3} \\
        One horizontal/vertical neighbor & $1$ & \motif{4} \\
        {One horizontal/vertical\\and one diagonal neighbor} & $1+2^{-3/2}=1.353\ldots$ & \motif{5} \\
        Two horizontal/vertical neighbors & $2$ & \motif{6} \\
    \end{tblr}
    \caption{The local motifs of spins that give rise to the peaks in the distribution of $J_i$, as seen in Fig.~3a of the main text. The value given in the center column is the local field on the central spin, which is the threshold of the corresponding feature in the distribution.
    \label{tab:Ji_motifs}}
\end{table}

In the limit $f \to 0$, the typical spacing is much greater than the lattice scale $r_\mathrm{typ} \gg a$, and the peaks associated to the structure of the lattice are lost. The distribution may be calculated exactly in this limit: we define the density of spins,
\begin{equation}\label{eq:rho}
    \rho = \frac{f}{a^2},
\end{equation}
with $a$ the lattice constant. The local field on a spin $i$ is given according to the distance $r_{\rm nn}$ to its nearest-neighbor by
\begin{equation}\label{eq:Ji}
    J_i = \int_{r_{\rm nn}}^\infty 2\pi r\rho\frac{J}{r^3} dr=\frac{2\pi \rho J}{r_{\rm nn}}.
\end{equation}
So we need the distribution of $r_{\rm nn}$. This can be calculated from the probability that an area $A$ is empty, $P_{\rm empty}(A)$. Subdividing that area into $n$ regions, in the limit $n\to\infty$, the probability of one such region being non-empty is $\rho \times A/n$. Thus,
\begin{equation}\label{eq:Pempty}
    P_{\rm empty}(A)=\lim_{n\to\infty}\left(1-\frac{\rho A}{n}\right)^n=e^{-\rho A}.
\end{equation}
The probability \emph{density} for $r_{\rm nn}$ is then,
\begin{equation}\label{eq:Pnn}
    p_{nn}(r_{\rm nn}) = \frac{d}{dr}\overbrace{\left(1-P_{\rm empty}\left(\pi r_{\rm nn}^2-\pi a^2\right)\right)}^\text{probability of spins within radius $r$}=2\pi r_{\rm nn} \rho\, e^{-\rho(\pi r_{\rm nn}^2-\pi a^2)}.
\end{equation}
Using the relation in Eq.~\eqref{eq:Ji}, this can be converted into a probability density for local fields:
\begin{equation}
    p_J(J_i)=p_{\rm nn}\left(r_{\rm nn}=\frac{2\pi\rho J}{J_i}\right)\left|\frac{dr_{\rm nn}}{dJ_i}\right|=\frac{8\pi^3\rho^3J^2}{J_i^3}e^{-\rho(4\pi^3\rho^2J^2/J_i^2-\pi a^2)}
\end{equation}

Note that $r_{\rm typ}$ is defined as the typical nearest-neighbor distance according to the distribution in Eq.~\eqref{eq:Pnn}. One common measure for this is the geometric mean, given by,
\begin{equation}\label{eq:rtyp_prop}
    r_{\rm typ} \propto e^{\langle \log (r_{\rm nn})\rangle_{p_{\rm nn}}}=e^{-\gamma/2}\frac{a}{\sqrt{\pi f}},
\end{equation}
with Euler's constant $\gamma = 0.577\ldots$. This confirms the expectation $r_{\rm typ}\sim 1/\sqrt{f}$.

\section{Cluster Discrete Truncated Wigner Approximation (cDTWA)}

To capture squeezing dynamics, we use the same cDTWA algorithm as in Ref.~\cite{wu_spin_2025}, which itself is adapted from Refs.~\cite{schachenmayer_manybody_2015a,braemer_cluster_2024}. All spins are clustered into pairs according to a real-space matching algorithm, which iteratively pairs and removes from consideration the two closest spins, continuing until all spins are paired. The interactions within clusters are treated quantum mechanically, whereas the interactions between clusters are treated at the mean-field level, with trajectories sampled over the discrete Wigner distribution corresponding to the initial $x$-state. This method captures squeezing dynamics effectively and can achieve large system sizes of up to $N \sim 10^4$ at low filling with disorder averaging.

\begin{figure}[hbt]
    \centering
    \includegraphics{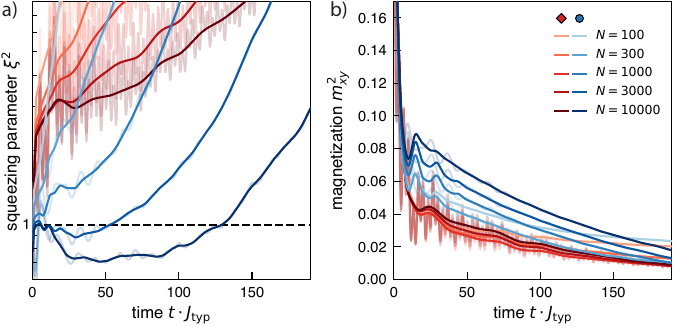}
    \caption{Dynamical evolution under cDTWA. (a) As in Fig.~3c of the main text, we show time evolution of $\xi^2$ at $f=0.05$, $\Delta=0$, when shelving $1.5\%$ (red, diamond) and $35\%$ (blue, circle) of spins. The solid curves are computed from the raw data (faded curves) by filtering out the dimer oscillations. (b) Magnetization $m_{xy}^2$ from the same data as panel (a).}
    \label{fig:dtwa}
\end{figure}

We use cDTWA to calculate the Wineland squeezing parameter, $\xi^2 = N {\rm{min}}_{\hat{n}\perp\hat{x}}{\rm{Var}}\left[\hat{n}\cdot\vec{S}\right]/\langle S^x \rangle^2$, with $\vec{S}=\sum_i \vec{s}_i$ the total spin operator. $\xi^2<1$ indicates metrological squeezing~\cite{wineland_squeezed_1994}. We can also compute the magnetization $m_{xy}^2$, which should thermalize to a non-zero finite value for $E_x < E_\mathrm{c}$, and decay to zero for $E_x > E_\mathrm{c}$. This is shown in Fig.~\ref{fig:dtwa}, comparing minimal and significant dimer shelving. The data in panel (a) corresponds to that of Fig.~3c of the main text.

Neither case in Fig.~\hyperref[fig:dtwa]{\ref*{fig:dtwa}b} shows a plateau in $m_{xy}^2$, since we only go to the early, prethermal times at which squeezing occurs. However, we do see that $m_{xy}^2$ grows with increasing system size only for the case of significant shelving such that $E_x < E_\mathrm{c}$ (see Fig.~3b in the main text), and it is expected to plateau at late times. On the other hand, for the case with minimal shelving such that $E_x > E_\mathrm{c}$, we find quick convergence in system size, and the converged curve is expected to decay to $0$.

At low filling fractions, on top of the squeezing dynamics, there are faster oscillations associated with dimers. For example, when prepared in the $x$-state, the closest dimers, which are a distance $1$ on the lattice, oscillate with frequency $\frac{1}{4\pi}$. The next-closest dimers, which are a distance $\sqrt{2}$ (diagonal neighbors) oscillate with frequency $\frac{1}{4\pi}\frac{1}{2^{3/2}}$, and so on. Applying notch filters at the fastest 3 dimer oscillations yields our smoothed curves.

\section{Mean-field theory and corrections from quantum correlations}

In this section, we derive the mean-field theory and several corrections that capture different leading-order quantum correlations. We work with the same Hamiltonian as $H_{\rm XXZ}$ [Eq.~(1) in the main text], but rewritten to absorb the anisotropy into the interaction matrix:
\begin{equation}\label{eq:H_XXZ}
    H = -\frac{1}{2}\sum_{ij}\sum_{a\in\{x,y,z\}}s_i^a J_{ij}^a s_j^a,\qquad J_{ij}^a=\left[1+(\Delta-1)\delta_{az}\right](1-\delta_{ij})J_{ij},\quad J_{ij} = Jr_{ij}^{-3}
\end{equation}
where the sum over $i,j$ is over \emph{spins} rather than lattice sites, of which only a fraction $f$ are populated.

The starting point for all these is the pure mean-field theory in Sec.~\ref{sec:MFT}, which does not capture any quantum fluctuations. As a consequence, it predicts $T_\mathrm{c} \sim f$ rather than $T_\mathrm{c} \sim f^{3/2}$, and it does not have any dependence on the Ising anisotropy $\Delta$.

To capture few-body quantum correlations, we perform a high-temperature cluster expansion, which is derived in Sec.~\ref{sec:CE}. In Sec.~\ref{sec:CE1}, we explicitly calculate $T_\mathrm{c}$ to 1st order with this expansion. This calculation does introduce $\Delta$-dependence, as well as a crossover between two different $f$ scalings for the strong and weak disorder limit. It turns out to therefore work well near $f=1$ but fail in the $f\to0$ limit, both because $T_\mathrm{c}$ still scales incorrectly with $f$ in this limit, and the location $\Delta_{\rm peak}$ of the peak in $T_\mathrm{c}$ does not shift towards the Heisenberg point.

In Sec.~\ref{sec:DMFT}, we consider a separate extension of the mean field ansatz by including the exact correlations of the strongest-interacting pairs (i.e.\ the ``dimers''). This correction works well for $f\to0$, as it numerically produces the proper scaling of $T_\mathrm{c}$ with $f$, and it successfully reproduces the shift in the peak of $T_\mathrm{c}$ towards the Heisenberg point.

Finally, in Sec.~\ref{sec:CE2}, we return to the cluster expansion and derive the 2nd order contribution. It turns out this is also enough to capture the proper shift in the peak of $T_\mathrm{c}$ as a function of $\Delta$, although it cannot be used to produce the full $T_\mathrm{c}$ curve.

\subsection{\label{sec:MFT}Mean-field theory}

Here we derive the mean-field estimate of $T_\mathrm{c}$ and show that it fails to capture the correct dependence on both $\Delta$ and $f$. The mean-field calculation makes use of a variational state $\rho_0$ given in terms of the total spin operator $S^a = \sum_i s_i^a$ as
\begin{equation}\label{eq:Z_0}
    \rho_0 = \frac{1}{Z_0}e^{-\beta H_0}, \qquad Z_0=\tr{e^{-\beta H_0}}, \qquad H_0 = -h S^x
\end{equation}
where $H_0$ is the variational Hamiltonian, and $Z_0$ the corresponding partition function. Expectation values with respect to this state are denoted $\langle O \rangle_0=\tr{\rho_0 O}$.

An upper bound to the true free energy is provided by the variational free energy
\begin{equation}
    F_0=\langle H\rangle_0-T S_0 
\end{equation}
where the entropy is given by the usual form $S_0 =-\tr{\rho_0 \ln(\rho_0)}$. We note the variational may be written as
\begin{equation}\label{eq:F_0}
    F_0 = \langle H_{\rm diff} \rangle_0 - T \ln(Z_0),\qquad\text{with}\qquad  H_{\rm diff}=H-H_0.
\end{equation}
The approximate phase diagram of the system may be found by minimizing $F_0$ with respect to the variational parameter $h$. A minimum at $h=0$ corresponds to the paramagnetic phase, while $|h|>0$ corresponds to long-range XY order. Thus the critical point $T_\mathrm{c} = 1/\beta_\mathrm{c}$ occurs when $h=0$ changes from a minimum to a maximum:
\begin{equation}\label{eq:Tc_condition_MF}
    \left.\frac{\partial^2 F_0}{\partial h^2}\right|_{h=0}=0.
\end{equation}

We will treat the two terms in Eq.~\eqref{eq:F_0} separately, and doing so requires us to introduce the zero-field magnetic susceptibility,
\begin{equation}\label{eq:chi_ij}
    \chi_{ij}^{ab} = \int_0^\beta d\tau\, \langle s_i^a(0) s_j^b(\tau)\rangle_{00} = \frac{\beta}{4}\delta_{ij} \delta_{ab},
\end{equation}
where we have introduced for an operator $O$ the notation, $\langle O \rangle_{00}=\langle O \rangle_0 |_{h=0}$, and where $O(\tau)=e^{-\tau H_0}Oe^{\tau H_0}$ is the Heisenberg-picture operator imaginary-time-evolved under $H_0$. As seen from the last equality, in the mean-field theory the susceptibility is fully diagonal.

This susceptibility has the relation,
\begin{align}
    \left.\frac{\partial \langle s_i^a \rangle_0}{\partial h}\right|_{h=0} = \int_0^\beta d\tau\, \langle  s_i^a(0)S^x(\tau)\rangle_{00} = \sum_j \int_0^\beta d\tau\, \langle s_i^a(0)s_j^x(\tau)\rangle_{00}&=\sum_j \chi_{ij}\delta_{a x}\label{eq:ds_dh}\\
    &\text{with}\ \chi_{ij} := \chi_{ij}^{xx}=\frac{\beta}{4}\delta_{ij}.\label{eq:chi}
\end{align}

Now the derivative of the first term in Eq.~\eqref{eq:F_0} is given by,
\begin{align}\label{eq:d_Hdiff}
    \left.\partial_h^2\langle H_{\rm diff} \rangle_0\right|_{h=0} &= -\frac{1}{2}\sum_{ij}\sum_{a\in\{x,y,z\}}\left.\partial_h^2\left(\left\langle s_i^a \right\rangle_0 J^a_{ij} \left\langle s^a_j\right\rangle_0\right)\right|_{h=0} + \left.\partial_h^2 \left(h\sum_i \langle s_i^x\rangle_0\right)\right|_{h=0}\notag\\
    &= -\sum_{ij}\sum_{a\in\{x,y,z\}}\left.\frac{\partial\langle s_i^a\rangle_0}{\partial h}\right|_{h=0} J^a_{ij} \left.\frac{\partial\langle s_j^a\rangle_0}{\partial h}\right|_{h=0} + 2\sum_i \left.\frac{\partial\langle s^x_i\rangle_0}{\partial h}\right|_{h=0}\notag\\
    &= -\sum_{ijkl}\, \chi_{ij}\, J_{jk}\chi_{kl} + 2\sum_{ij}\chi_{ij}.
\end{align}
where for the second line we have kept only the surviving terms of the product rule according to $\langle s^a_i \rangle_{00}=0$ by the symmetry $s_i^a\to-s_i^a$ of $H$ [Eq.~\eqref{eq:H_XXZ}] at $h=0$.

We also have,
\begin{align}\label{eq:d_Z}
    \left.\partial_h^2(-T \ln Z_0)\right|_{h=0} &= -T\partial_h \left.\left(Z_0^{-1}\partial_h Z_0\right)\right|_{h=0}\notag\\
    &= -T\partial_h\left.\int_0^\beta d\tau\, \langle S^x(\tau)\rangle_0\right|_{h=0}\notag\\
    &= -\sum_i\left.\frac{\partial \langle s_i^x\rangle_0}{\partial h}\right|_{h=0} = -\sum_{ij} \chi_{ij}.
\end{align}
where for the third line we have used the fact that $\langle O(\tau)\rangle = \langle O\rangle$ for any operator $O$.

Adding Eqs.~\eqref{eq:d_Hdiff} and \eqref{eq:d_Z}, at the critical temperature we have,
\begin{equation}\label{eq:Tc_MF_condition}
    \left.\frac{\partial^2 F_0}{\partial h^2}\right|_{h=0}=-\sum_{ijkl} \chi_{ij}\, J_{jk}\, \chi_{kl} + \sum_{ij}\chi_{ij} = 0 
\end{equation}

Substituting the form of $\chi_{ij}$ in Eq.~\eqref{eq:chi} yields the critical temperature,
\begin{equation}\label{eq:Tc_MF_1}
    T_\mathrm{c}^{\rm MF} = \frac{1}{4}\frac{1}{N}\sum_{ij}J_{ij}
\end{equation}
Again, we are implicitly summing over spins rather than lattice sites, which can be converted to a sum over all the lattice sites (defined as the set $\Lambda$) in the infinite system, to evaluate it directly. Defining $f_i$ as $1$ for a filled site and $0$ for an empty site,
\begin{align}\label{eq:Jij_sum}
    \frac{1}{N}\sum_{ij}J_{ij} &= \frac{1}{fN_\Lambda}\sum_{ij\in \Lambda}f_if_jJ_{ij} &&(N_\Lambda=L^2)\notag\\
    &= \frac{1}{f}\left\langle f_i\sum_{j\in \Lambda}f_jJ_{ij}\right\rangle_{i\in\Lambda} = \frac{1}{f}f\left\langle N_\Lambda\left\langle f_jJ_{ij}\right\rangle_{j\in\Lambda} \right\rangle_{i\in\Lambda}&&(\ \langle f_i\rangle_{i\in\Lambda}=f\ )\notag\\
    &= N_\Lambda \left\langle \left\langle f_j\right\rangle_{j\in\Lambda} \left\langle J_{ij}\right\rangle_{j\in\Lambda}\right\rangle_{i\in\Lambda}= N_\Lambda f\langle J_{ij} \rangle_{ij\in\Lambda}\notag\\
    &=f\left\langle \sum_{j\in\Lambda} J_{ij}\right\rangle_{i\in\Lambda},
\end{align}
where we have used the site-independence of $f_i$ to factor the averages over lattice sites.

The latter sum over lattice sites can be written exactly for any $i\in\Lambda$ as
\begin{equation}\label{eq:Jij_sum_numeric}
    \sum_{j\in\Lambda} J_{ij}=4 J\sum_{n=1}^\infty\sum_{m=0}^\infty(n^2+m^2)^{-3/2}=9.03\ldots
\end{equation}
Plugging this sum into Eq.~\eqref{eq:Tc_MF_1} gives us the mean-field critical temperature,
\begin{equation}\label{eq:Tc_MF_2}
    T_\mathrm{c}^{\rm MF} \approx 2.25\ldots\times J f.
\end{equation}

This estimate of $T_\mathrm{c}$ fails to produce any $\Delta$ dependence and scales as $\sim f$ rather than $\sim f^{3/2}$, so corrections are needed.

\subsection{\label{sec:CE}High temperature cluster expansion}

In this section, we lay out the formalism for our high-temperature cluster expansion, which captures corrections to the mean-field theory from few-body correlations. Following the introduce here, in Sec.~\ref{sec:CE1} and Sec.~\ref{sec:CE2} we derive the full 1st order and 2nd order corrections, respectively.

The mean field critical temperature is large compared to $J_{\rm typ}$, so as a correction to the mean-field treatment that includes quantum correlations, we perform a high-temperature (cumulant) cluster expansion~\cite{kubo_generalized_1962,Oitmaa_Hamer_Zheng_2006} of the variational free energy in powers of $\beta$. We first write the new variational ansatz,
\begin{equation}\label{eq:H_lambda}
    H_\lambda = H_0 + \lambda H_{\rm diff}
\end{equation}
where we introduce an expansion parameter $0\le\lambda\le 1$ which allows us to tune between the mean-field Hamiltonian $H_0$ and the true Hamiltonian $H_1=H$. In particular, we want to perturbatively reintroduce the correlations in $H$ that were neglected in $H_0$. Motivated by the observation that $\beta_\mathrm{c} H \ll 1$, we do this by expanding in powers of $\lambda$, truncating (since each power of $\lambda$ comes with a power of $\beta H$), and taking $\lambda \to 1$. We use the subscript $\lambda$ in place of the subscript $0$ to indicate the new ansatz.

Using the identity $\partial_\lambda\ln(Z_\lambda)=-\beta\langle H_{\rm diff}\rangle_\lambda$, we obtain the variational free energy,
\begin{equation}\label{eq:F_lambda}
    F_\lambda = \langle H - H_\lambda \rangle_\lambda - T \ln(Z_\lambda) = -T\left[(1-\lambda)\partial_\lambda\ln(Z_\lambda)+\ln(Z_\lambda)\right].
\end{equation}

We make use of the time-ordered expansion,
\begin{equation*}
    e^{-\beta H_\lambda}=\sum_{n=0}^\infty \frac{(-\lambda)^n}{n!}\int_0^\beta d\tau_1\cdots\int_0^\beta d\tau_n \mathcal{T}\left[H_{\rm diff}(\tau_1)\cdots H_{\rm diff}(\tau_n)\right]e^{-\beta H_0}
\end{equation*}
where $\mathcal{T}$ indicates time ordering and the Heisenberg picture time evolution is still under $H_0$. It allows us to write,
\begin{equation}\label{eq:correlators}
    \frac{Z_\lambda}{Z_0}=\sum_{n=0}^\infty \frac{(-\lambda)^n}{n!} C_n\quad\text{with}\quad C_n = \int_0^\beta d\tau_1\cdots\int_0^\beta d\tau_n \left\langle\mathcal{T}\left[H_{\rm diff}(\tau_1)\cdots H_{\rm diff}(\tau_n)\right]\right\rangle_0.
\end{equation}
where the moments $C_n$ are the n-point correlators of $H_{\rm diff}$.

Then $\ln(Z_\lambda)$ can be written as,
\begin{equation}\label{eq:cumulants}
   \ln\left(\frac{Z_\lambda}{Z_0}\right)=\sum_{n=1}^\infty \frac{(-\lambda)^n}{n!}K_n,\quad\text{with}\quad K_n = \sum_{k=0}^n (-1)^{k-1} (k-1)!\, \mathcal{B}_{n,k}(C_1,\cdots,C_{n-k+1}),
\end{equation}
where $K_n$ are the corresponding cumulants, i.e.\ $n$-point connected correlators, written here as a function of the moments $\{C_n\}$ via the incomplete Bell polynomials $\mathcal{B}_{n,k}$. We will only make use of up to $n=3$, so the cumulants we care about are given by,
\begin{equation}\label{eq:bell_polynomials}
    K_0 = 0,\quad K_1 = C_1,\quad K_2 = C_2 - C_1^2,\quad K_3 = C_3 - 3C_2C_1 + 2C_1^3.
\end{equation}

Substituting into Eq.~\eqref{eq:F_lambda}, we arrive at the free energy expansion in powers of $\lambda$, which gives the order $n$ approximation,
\begin{equation}\label{eq:F_expansion}
    F^{(n)}_\lambda = \sum_{m=0}^n\frac{(-\lambda)^m}{m!}F_m\quad\text{with}\quad F_m=\begin{cases}
        T(K_1-\ln Z_0) & m=0\\
        T(K_{m+1} + (m-1)K_m) & m>0
    \end{cases}
\end{equation}
where $F_0$ is just the mean-field free energy (Eq.~\eqref{eq:F_0}).

Calculating the critical temperature requires us to compute the second derivative $F''_\lambda$, which means we need to be able to take derivatives of any given correlator $C_n$. The correlators and their derivatives can all be expressed in terms of the object,
\begin{equation}\label{eq:A_nm}
    A_{n,m}=\frac{1}{D}\int_0^\beta d\tau_1\cdots\int_0^\beta d\tau_{n+m}\tr{\mathcal{T}\left[H_{\rm diff}(\tau_1)\cdots H_{\rm diff}(\tau_n)H_{\rm diff}'(\tau_{n+1})\cdots H_{\rm diff}'(\tau_{n+m})\right]e^{-\beta H_0}},
\end{equation}
with $D=2^N$ the Hilbert space dimension, and noting that $H_{\rm diff}'=-H_0'=S^x$ since $H$ is $h$-independent. Indeed, this object's derivative with respect to $h$ has the recursion relation,
\begin{equation}\label{eq:A_derivative_recurrence}
    A_{n,m}'=nA_{n-1,m+1}+A_{n,m+1},
\end{equation}
which can be seen with the following counting arguments from the terms of the product rule.

If the derivative hits a $H_{\rm diff}(\tau)$, it becomes $H_{\rm diff}'(\tau)$, so $n \to n-1$ and $m \to m+1$ with $n$ such terms. This is the first term of Eq.~\eqref{eq:A_derivative_recurrence}. If on the other hand the derivative hits a $H_{\rm diff}'(\tau)$, it vanishes (since $H_0'=-S^x$ has no $h$ dependence). Finally if the derivative hits the time evolution, we use the integral identity
\begin{equation}\label{eq:derivative_of_time_evolution}
    \partial_h e^{-\tau H_0}=\int_0^\tau d\tau' e^{-\tau' H_0}(-H_0')e^{-(\tau-\tau')H_0}=\int_0^\tau d\tau' H_{\rm diff}'(\tau')e^{-\tau H_0}
\end{equation}
to see that an extra $H_{\rm diff}'(\tau)$ is introduced with the appropriate time-ordering, so $m \to m+1$. This is the second term of Eq.~\eqref{eq:A_derivative_recurrence}.

We can calculate higher order derivatives by considering the paths through which the recurrence relation of Eq.~\eqref{eq:A_derivative_recurrence} traverses the $\{n,m\}$ lattice of the $A_{n,m}$ objects. The $k$th derivative takes $m\to m+k$ and $n\to n-p$ with $p$ ranging from $0$ to $k$. For a given $p$, there are $\binom{k}{p}$ paths to get from $(n,m)$ to $(n-p,m+k)$ with successive differentiation, which correspond to which $p$ of the $k$ steps contribute the first term in Eq.~\eqref{eq:A_derivative_recurrence}. Each step carries a factor of its current $n$, which totals to a factor of $n!/(n-p)!=p!\binom{n}{p}$. This gives the general formula for the $k$th derivative,
\begin{equation}\label{eq:A_derivatives}
    \frac{\partial^k A_{n,m}}{\partial h^k}=\sum_{p=0}^k p! \binom{k}{p}\binom{n}{p}A_{n-p,m+k}.
\end{equation}

Since we will want to set $h=0$ after taking derivatives, it will be helpful to work with the Taylor series expansion around $h=0$,
\begin{equation}\label{eq:A_taylor}
    A_{n,m}=\sum_{k=0}^\infty \frac{h^k}{k!}A_{n,m}^{(k)}\quad\text{with}\quad A_{n,m}^{(k)} := \left.\frac{\partial^k A_{n,m}}{\partial h^k}\right|_{h=0}=\sum_{p=0}^k p! \binom{k}{p}\binom{n}{p}A^{(0)}_{n-p,m+k},
\end{equation}
where a superscript ${(k)}$ indicates taking the $k$th derivative and then setting $h=0$, and therefore that we have implicitly defined $A_{n,m}^{(0)}=A_{n,m}|_{h=0}$.

We can relate the correlators of Eq.~\eqref{eq:correlators} to these objects,
\begin{equation}\label{eq:C_from_A}
    C_n=\frac{A_{n,0}}{A_{0,0}}=\frac{\sum_{k=0}^\infty \frac{h^k}{k!}\sum_{p=0}^k p!\binom{k}{p}\binom{n}{p}A^{(0)}_{n-p,k}}{\sum_{k=0}^\infty \frac{h^k}{k!}A^{(0)}_{0,k}},
\end{equation}
so the $h$ dependence is explicit.

We now just need to be able to compute $A_{n,m}^{(0)} = A_{n,m}|_{h=0}$. Returning to the definition of $A_{n,m}$ in Eq.~\eqref{eq:A_nm}, since $H_0|_{h=0}=0$, we have $H_{\rm diff}(\tau)|_{h=0}=H_{\rm diff}|_{h=0}=H$, as well as $H_{\rm diff}'(\tau)|_{h=0}=H_{\rm diff}'|_{h=0}=S^x$. Despite all operators remaining constant in imaginary time, their time-ordering still matters since they do not commute, and we have,
\begin{equation}\label{eq:A_nm_0}
    A_{n,m}^{(0)}=\frac{1}{D}\beta^{n+m}\tr{\overline{H^n(S^x)^m}},
\end{equation}
where the overline indicates an average over all permutations of the ordering of the $n+m$ operators (though many are equivalent because of the trace).

We can immediately calculate some of its values. If $m$ is odd, $A_{n,m}^{(0)}=0$, since the product inside the trace is off-diagonal in the $S^z$ basis (conserved by $H$) and thus the trace vanishes. We also have $A^{(0)}_{0,0}=1$, and $A^{(0)}_{1,0}=0$ since $H$ is traceless. 

Applying these rules to the derivatives of $C_n$ in Eq.~\eqref{eq:C_from_A}, we have,
\begin{subequations}\label{eq:C_ds}
\begin{align}
    C_n|_{h=0}&=A_{n,0}^{(0)},\label{eq:C_d0}\\
    C'_n|_{h=0}&=0,\label{eq:C_d1}\\
    C''_n|_{h=0}&=A^{(0)}_{n,2}-A^{(0)}_{n,0}A^{(0)}_{0,2}+2nA^{(0)}_{n-1,2}+n(n-1)A^{(0)}_{n-2,2}.\label{eq:C_d2}
\end{align}
\end{subequations}

\subsection{\label{sec:CE1}1st order cluster expansion}

We now show that to 1st order in $\lambda$, the cluster expansion around the mean-field $T_\mathrm{c}$ introduces $\Delta$ dependence and is successful near $f=1$. Combining Eqs. \ref{eq:bell_polynomials} and \ref{eq:C_ds}, we have the (intensive, dimensionless) quantities,
\begin{subequations}\label{eq:K_ds}
\begin{align}
    \frac{1}{\beta^2N}K''_1|_{h=0}&=\beta B_1+2 B_0,\label{eq:K1_d}\\
    \frac{1}{\beta^2N}K''_2|_{h=0}&=\beta^2 B_2+4 \beta B_1+2 B_0,\label{eq:K2_d}
\end{align}
\end{subequations}
where we have defined the fully-connected parts of the objects in Eq.~\eqref{eq:A_nm_0} and rescaled to make them $\beta$-independent and intensive,
\begin{subequations}\label{eq:B}
\begin{align}
    B_0&=N^{-1}\beta^{-2}A_{0,2}^{(0)}&&=\frac{1}{4},\label{eq:B0}\\
    B_1&=N^{-1}\beta^{-3}A_{1,2}^{(0)}&&=-\frac{1}{16}f\frac{1}{L^2}\sum_{ij\in\Lambda}J_{ij},\label{eq:B1}\\
    B_2&=N^{-1}\beta^{-4}\left(A_{2,2}^{(0)}-A_{2,0}^{(0)}A_{0,2}^{(0)}\right)&&=\frac{1}{64}\left(2f^2\frac{1}{L^2}\sum_{ijk\in\Lambda}J_{ij}J_{jk}-\frac{2}{3}(4+\Delta+\Delta^2)f\frac{1}{L^2}\sum_{ij\in\Lambda}J_{ij}^2\right).\label{eq:B2}
\end{align}
\end{subequations}

The final expressions in Eqs.~\eqref{eq:B} were obtained directly from Eq.~\eqref{eq:A_nm_0} with a sum over fully-connected diagrams, which is explained in the remainder of this section. $H$ and $S^x$ are sums over spin operators, and the only terms that have a nonzero trace are those in which all operators act on spins in pairs with the same spin component or in triplets with one of each spin component (e.g.\ $\tr{s_i^x s_i^x}$ or $\tr{s_i^x s_i^ys_i^z}$). For the diagrammatic language, we represent each single site operator as a dot, of which there are $2n+m$ for $A_{n,m}^{(0)}$. If one operator is indexed $i$ and another is indexed $j$, there is a Kronecker delta function $\delta_{ij}$ to indicate that these operators must act on the same site, which we represent as a black line between them. If two operators come from $H$, they also cannot act on the same site, which is shown by an ellipse encompassing them. The order of the operators matters and is cyclic due to the trace. 

We use two additional line styles to denote necessary conditions to keep only the fully-connected diagrams. These are the diagrams which cannot separate into an equivalent product of subdiagrams, due to sign flips from commutation rules. We represent $(1-\delta_{ij})$ as a red jagged line---necessary for diagrams where those operators acting on the same site would change the commutation rules such that it becomes one of the disconnected diagrams. On the other hand, the situation can also arise where several operators acting on the same site and on different spin components would make an otherwise disconnected diagram a connected one due to commutation rules. In these scenarios we use double lines to show that two pairs act on the same sites, in which case we must first subtract off the corresponding disconnected diagram that is not canceled, and then add the contribution from the connected one. (We similarly use triple lines to show another pair acting on the same site as each other, but distinct from the single or double lines.)

The simplest diagram that arises corresponds to :
\begin{align*}
    B_0=\beta^{-2} A_{0,2}^{(0)}=D^{-1}\tr{S^xS^x}&=D^{-1}\sum_{nm}\tr{s_n^xs_m^x}\\
    &=
    \begin{diagram}{2}
        \connect{1}{2}
    \end{diagram}
    =\frac{1}{4}\sum_{nm}\delta_{nm}=\frac{1}{4}N.
\end{align*}

When there are more than two operators, we must count the different ways they can pair up:
\renewcommand{\arraystretch}{4}
\begin{align*}
    B_1=\beta^{-3} A_{1,2}^{(0)}=D^{-1}\tr{HS^xS^x}&=-D^{-1}\frac{1}{2}\sum_{ij}\sum_{\alpha\in\{x,y,z\}}\sum_{nm}J_{ij}^\alpha\tr{s_i^\alpha s_j^\alpha s_n^x s_m^x}\\
    &= 
    \begin{diagram}{4}
        \encircle{2}{3}
        \connect{1}{2}
        \connect{3}{4}
    \end{diagram}
    +
    \begin{diagram}{4}
        \encircle{2}{3}
        \connect{1}{3}
        \connect{2}{4}
    \end{diagram}\\
    &=-\frac{1}{16}\frac{1}{2}\sum_{ij}\sum_{\alpha\in\{x,y,z\}}\sum_{nm}J_{ij}^\alpha(\delta_{in}\delta_{jm}\delta_{\alpha x}+\delta_{im}\delta_{jn}\delta_{\alpha x})=-\frac{1}{16}\sum_{ij}J_{ij}.
\end{align*}

Finally we turn to $B_2$, for which we directly enumerate the relevant diagrams. In the corresponding equations for each set of diagrams, we implicitly sum over all indices present, with Latin letters for spin indices and Greek letters for Pauli indices. Note that there are two inequivalent orderings of the operators in the trace: $\tr{\overline{HHS^sS^x}}=\frac{2}{3}\tr{H H S^x S^x}+\frac{1}{3}\tr{H S^x H S^x}$. The corresponding diagrams for the first term are given in Table~\ref{tab:HHXX}, while those for the second are given in Table~\ref{tab:HXHX}. Summing over the appropriately weighted contributions from all the diagrams gives the final expression for $B_2$ shown in Eq.~\eqref{eq:B2}.

\begin{table}[h]
\centering
\begin{tabular}{c c c c | c}
    $\begin{diagram}{6}
        \encircle{2}{3}
        \encircle{4}{5}
        \disconnect{1}{6}
        \connect{1}{2}
        \connect{6}{5}
        \connect{3}{4}
    \end{diagram}$ &
    $\begin{diagram}{6}
        \encircle{2}{3}
        \encircle{4}{5}
        \disconnect{1}{6}
        \connect{1}{2}
        \connect{6}{4}
        \connect{3}{5}
    \end{diagram}$ &
    $\begin{diagram}{6}
        \encircle{2}{3}
        \encircle{4}{5}
        \disconnect{1}{6}
        \connect{1}{3}
        \connect{6}{5}
        \connect{2}{4}
    \end{diagram}$ &
    $\begin{diagram}{6}
        \encircle{2}{3}
        \encircle{4}{5}
        \disconnect{1}{6}
        \connect{1}{3}
        \connect{6}{4}
        \connect{2}{5}
    \end{diagram}$
    & \multirow[c]{2}{*}{\rule{0pt}{1.5cm}
    \(\displaystyle 
    \begin{aligned}
       &D^{-1}\frac{1}{4}\sum J_{ij}^\alpha J_{kl}^\beta\frac{1}{64} \delta_{in}\delta_{jk}\delta_{ml}\delta_{\alpha x}\delta_{\beta x}(1-\delta_{nm})\\
        &= \frac{1}{64}\frac{1}{4}\left(\sum_{ijk} J_{ij} J_{jk}-\sum_{ij}J_{ij}^2\right)
    \end{aligned}
    \)
    } \\
    $\begin{diagram}{6}
        \encircle{2}{3}
        \encircle{4}{5}
        \disconnect{1}{6}
        \connect{1}{5}
        \connect{6}{2}
        \connect{3}{4}
    \end{diagram}$ &
    $\begin{diagram}{6}
        \encircle{2}{3}
        \encircle{4}{5}
        \disconnect{1}{6}
        \connect{1}{4}
        \connect{6}{2}
        \connect{3}{5}
    \end{diagram}$ &
    $\begin{diagram}{6}
        \encircle{2}{3}
        \encircle{4}{5}
        \disconnect{1}{6}
        \connect{1}{5}
        \connect{6}{3}
        \connect{2}{4}
    \end{diagram}$ &
    $\begin{diagram}{6}
        \encircle{2}{3}
        \encircle{4}{5}
        \disconnect{1}{6}
        \connect{1}{4}
        \connect{6}{3}
        \connect{2}{5}
    \end{diagram}$ & \\
    [3ex]\hline

    $\begin{diagram}{6}
        \encircle{2}{3}
        \encircle{4}{5}
        \connect{1}{2}
        \connect{2}{4}
        \connect{4}{1}
        \connect{6}{5}
        \connect{5}{3}
        \connect{3}{6}
    \end{diagram}$ &
    $\begin{diagram}{6}
        \encircle{2}{3}
        \encircle{4}{5}
        \connect{1}{2}
        \connect{2}{5}
        \connect{5}{1}
        \connect{6}{4}
        \connect{4}{3}
        \connect{3}{6}
    \end{diagram}$ &
    $\begin{diagram}{6}
        \encircle{2}{3}
        \encircle{4}{5}
        \connect{1}{3}
        \connect{3}{4}
        \connect{4}{1}
        \connect{6}{5}
        \connect{5}{2}
        \connect{2}{6}
    \end{diagram}$ &
    $\begin{diagram}{6}
        \encircle{2}{3}
        \encircle{4}{5}
        \connect{1}{3}
        \connect{3}{5}
        \connect{5}{1}
        \connect{6}{4}
        \connect{4}{2}
        \connect{2}{6}
    \end{diagram}$
    &
    \(\displaystyle 
    \begin{aligned}
       &D^{-1}\frac{1}{4}\sum J_{ij}^\alpha J_{kl}^\beta\frac{1}{64} \delta_{in}\delta_{kn}\delta_{jm}\delta_{lm}(\delta_{\alpha y}\delta_{\beta z}+\delta_{\alpha z}\delta_{\beta y})\\
       & = \frac{1}{64}\frac{1}{4}\left(i^2+(-i)^2\right)\Delta\sum_{ij}J_{ij}^2=-\frac{1}{64}\frac{1}{2}\Delta\sum_{ij}J_{ij}^2
    \end{aligned}
    \)
\end{tabular}
\caption{Diagrams and their individual evaluation that correspond to the fully-connected contributions to $\tr{HHS^xS^x}$ as appears in $A^{(0)}_{2,2}$~\eqref{eq:A_nm_0}}\label{tab:HHXX}
\end{table}

\begin{table}[h]
\centering
\begin{tabular}{c c c c | c}
    $\begin{diagram}{6}
        \encircle{6}{1}
        \encircle{3}{4}
        \disconnect{2}{5}
        \connect{5}{6}
        \connect{2}{3}
        \connect{1}{4}
    \end{diagram}$ &
    $\begin{diagram}{6}
        \encircle{6}{1}
        \encircle{3}{4}
        \disconnect{2}{5}
        \connect{5}{1}
        \connect{2}{3}
        \connect{6}{4}
    \end{diagram}$ &
    $\begin{diagram}{6}
        \encircle{6}{1}
        \encircle{3}{4}
        \disconnect{2}{5}
        \connect{5}{6}
        \connect{2}{4}
        \connect{1}{3}
    \end{diagram}$ &
    $\begin{diagram}{6}
        \encircle{6}{1}
        \encircle{3}{4}
        \disconnect{2}{5}
        \connect{5}{1}
        \connect{2}{4}
        \connect{6}{3}
    \end{diagram}$
    & \multirow[c]{2}{*}{\rule{0pt}{1.5cm}
    \(\displaystyle 
    \begin{aligned}
       &D^{-1}\frac{1}{2}\sum J_{ij}^\alpha J_{kl}^\beta\frac{1}{64} \delta_{in}\delta_{jk}\delta_{ml}\delta_{\alpha x}\delta_{\beta x}(1-\delta_{nm})\\
        &= \frac{1}{64}\frac{1}{4}\left(\sum_{ijk} J_{ij} J_{jk}-\sum_{ij}J_{ij}^2\right)
    \end{aligned}
    \)
    } \\
    $\begin{diagram}{6}
        \encircle{6}{1}
        \encircle{3}{4}
        \disconnect{2}{5}
        \connect{2}{1}
        \connect{5}{4}
        \connect{6}{3}
    \end{diagram}$ & 
    $\begin{diagram}{6}
        \encircle{6}{1}
        \encircle{3}{4}
        \disconnect{2}{5}
        \connect{2}{6}
        \connect{5}{4}
        \connect{1}{3}
    \end{diagram}$ & 
    $\begin{diagram}{6}
        \encircle{6}{1}
        \encircle{3}{4}
        \disconnect{2}{5}
        \connect{2}{1}
        \connect{5}{3}
        \connect{6}{4}
    \end{diagram}$ & 
    $\begin{diagram}{6}
        \encircle{6}{1}
        \encircle{3}{4}
        \disconnect{2}{5}
        \connect{2}{6}
        \connect{5}{3}
        \connect{1}{4}
    \end{diagram}$ &
    \\
    [3ex]\hline
    
    $\begin{diagram}{6}
        \encircle{6}{1}
        \encircle{3}{4}
        \connect[2]{2}{5}
        \connect{1}{3}
        \connect[2]{6}{4}
    \end{diagram}$ & 
    $\begin{diagram}{6}
        \encircle{6}{1}
        \encircle{3}{4}
        \connect[2]{2}{5}
        \connect{1}{4}
        \connect[2]{6}{3}
    \end{diagram}$ &
    $\begin{diagram}{6}
        \encircle{6}{1}
        \encircle{3}{4}
        \connect[2]{2}{5}
        \connect[2]{1}{3}
        \connect{6}{4}
    \end{diagram}$ & 
    $\begin{diagram}{6}
        \encircle{6}{1}
        \encircle{3}{4}
        \connect[2]{2}{5}
        \connect[2]{1}{4}
        \connect{6}{3}
    \end{diagram}$ &
    \(\displaystyle 
    \begin{aligned}
       &D^{-1}\frac{1}{4}\sum J_{ij}^\alpha J_{kl}^\beta\frac{1}{64} \delta_{ik}\delta_{jl}\delta_{nm}\delta_{in}\delta_{\alpha \beta}\\
       &\qquad\qquad\qquad\times(-(\delta_{\alpha x}+\delta_{\alpha y}+\delta_{\alpha z})+(\delta_{\alpha x}-\delta_{\alpha y}-\delta_{\alpha z}))\\
        &= \frac{1}{64}\frac{1}{4}(-2-2\Delta^2)\sum_{ij}J_{ij}^2=-\frac{1}{64}\frac{1}{2}(1+\Delta^2)\sum_{ij}J_{ij}^2
    \end{aligned}
    \)
    \\
    [3ex]\hline
    
    $\begin{diagram}{6}
        \encircle{6}{1}
        \encircle{3}{4}
        \connect{2}{1}
        \connect{1}{3}
        \connect{3}{2}
        \connect{5}{6}
        \connect{6}{4}
        \connect{4}{5}
    \end{diagram}$ & 
    $\begin{diagram}{6}
        \encircle{6}{1}
        \encircle{3}{4}
        \connect{2}{1}
        \connect{1}{4}
        \connect{4}{2}
        \connect{5}{6}
        \connect{6}{3}
        \connect{3}{5}
    \end{diagram}$ & 
    $\begin{diagram}{6}
        \encircle{6}{1}
        \encircle{3}{4}
        \connect{2}{6}
        \connect{6}{3}
        \connect{3}{2}
        \connect{5}{1}
        \connect{1}{4}
        \connect{4}{5}
    \end{diagram}$ & 
    $\begin{diagram}{6}
        \encircle{6}{1}
        \encircle{3}{4}
        \connect{2}{6}
        \connect{6}{4}
        \connect{4}{2}
        \connect{5}{1}
        \connect{1}{3}
        \connect{3}{5}
    \end{diagram}$ &
    \(\displaystyle 
    \begin{aligned}
       &D^{-1}\frac{1}{2}\sum J_{ij}^\alpha J_{kl}^\beta\frac{1}{64} \delta_{in}\delta_{kn}\delta_{jm}\delta_{lm}(\delta_{\alpha y}\delta_{\beta z}+\delta_{\alpha z}\delta_{\beta y})\\
        &= \frac{1}{64}\frac{1}{4}\left(i(-i)+(-i)i\right)\Delta\sum_{ij}J_{ij}^2=\frac{1}{64}\frac{1}{2} \Delta\sum_{ij}J_{ij}^2
    \end{aligned}
    \)
\end{tabular}
\caption{Diagrams for $\tr{HS^xHS^x}$, shown in the same way as in Table ~\ref{tab:HHXX}.}\label{tab:HXHX}
\end{table}

Following the logic of Eq.~\eqref{eq:Jij_sum}, each sum over spins of products of $J_{ij}$ interactions can be converted to a sum over lattices sites $\Lambda$ instead, with the only change being $n-1$ factors of filling $f$ for a sum over $n$ indices.

From Eq.~\eqref{eq:F_expansion}, the free energy is given to 1st order in the cluster expansion by,
\begin{equation}\label{eq:f_1}
    F^{(1)}_\lambda=T(K_1 - \ln Z_0 - \lambda K_2).
\end{equation}
Using Eqs.~\eqref{eq:d_Z} and \eqref{eq:K_ds} to calculate the free energy per site $f=F/N$ to 1st order in $\lambda$, its second derivative at $h=0$ is,
\begin{align}\label{eq:f_1_d2}
    \left.\frac{\partial^2f^{(1)}_\lambda}{\partial h^2}\right|_{h=0}&=-\overbrace{\beta/4}^{\mathclap{T\ln Z_0}} + \overbrace{\beta^2 B_1+2\beta B_0}^{TK_1} - \lambda (\overbrace{\beta^3 B_2 + 4\beta^2 B_1 + 2 \beta B_0}^{TK_2})\notag\\
    &= \beta/4+\beta^2 B_1 - \lambda (\beta/2+4\beta^2 B_1 + \beta^3 B_2).
\end{align}
where the overbraces indicate the corresponding term in $F^{(1)}_\lambda$ from which each contribution is derived, and for the last line we have substituted Eq.~\eqref{eq:B0} for $B_0$

Setting Eq.~\eqref{eq:f_1_d2} equal to $0$ allows us to solve for the inverse critical temperature $\beta_\mathrm{c}$ in orders of $\lambda$. We find,
\begin{equation*}
    \beta_\mathrm{c}^{(\lambda)}=-\frac{1}{4B_1}+\lambda\frac{-8 B_1^2+B_2}{16B_1^3}+O(\lambda^2).
\end{equation*}
Truncating to $O(\lambda)$ and then taking $\lambda\to1$, we obtain the 1st order cluster expansion (CE1) correction to the mean-field inverse critical temperature $\beta_\mathrm{c}^{\rm MF}=(T_\mathrm{c}^{\rm MF})^{-1}$,
\begin{align*}
    \beta_\mathrm{c}^{\rm CE1}&=\left(3-\frac{2\frac{1}{L^2}\sum_\Lambda J_{ij}J_{jk}-\frac{2}{3}(4+\Delta+\Delta^2)f^{-1}\frac{1}{L^2}\sum_\Lambda J_{ij}^2}{\left(\frac{1}{L^2}\sum_\Lambda J_{ij}\right)^2}
    \right)\beta_\mathrm{c}^{\rm MF}\\
    &= \left(1+\frac{2(4+\Delta+\Delta^2)f^{-1}\frac{1}{L^2}\sum_\Lambda J_{ij}^2}{3\left(\frac{1}{L^2}\sum_\Lambda J_{ij}\right)^2}
    \right)\beta_\mathrm{c}^{\rm MF}
\end{align*}
where $\sum_\Lambda$ is implicitly for all the indices over all the lattice sites. For the second line, we have used the fact that
\begin{equation*}
    \frac{1}{L^2}\sum_{ijk\in\Lambda}J_{ij}J_{jk}=\frac{1}{L^2}\sum_{j\in\Lambda}\left(\sum_{i\in\Lambda}J_{ij}\right)^2=\left\langle\left(\sum_{i\in\Lambda}J_{ij}\right)^2\right\rangle_{i\in\Lambda}=\left\langle\sum_{i\in\Lambda}J_{ij}\right\rangle_{i\in\Lambda}^2=\left(\frac{1}{L^2}\sum_{ij\in\Lambda}J_{ij}\right)^2
\end{equation*}
since in the infinite system these sums are independent of the reference site being averaged over. We can also calculate $L^{-2}\sum_\Lambda J_{ij}^2$ by the same approach as Eq.~\eqref{eq:Jij_sum_numeric}, since for all $i\in\Lambda$,
\begin{equation}\label{eq:Jij_2_sum}
    \sum_{j\in\Lambda} J_{ij}^2=4\sum_{n=1}^\infty\sum_{m=0}^\infty
    \left((n^2+m^2)^{-3/2}\right)^2=4.658\ldots.
\end{equation}

We therefore arrive at the numeric result for the critical temperature,
\begin{equation}\label{eq:Tc_CE1}
    T_\mathrm{c}^{\rm CE1}=\left(1+0.0381\ldots\times\frac{4+\Delta+\Delta^2}{f}
    \right)^{-1}T_\mathrm{c}^{\rm MF}
\end{equation}
Notice that the only $\Delta$ dependence is the factor $4+\Delta+\Delta^2$, which is $f$-independent and extremized at $\Delta_{\rm peak}^{\rm CE1}=-0.5$.

We also note that $T_\mathrm{c}^{\rm CE1}$ has different behavior in $f$ depending on which term in the parentheses dominates. For the weak disorder limit ($f\approx 1$), the first term dominates and $T_\mathrm{c}^{\rm CE1}\sim T_\mathrm{c}^{\rm MF}\sim f$, while for the strong disorder limit ($f \ll 1$), the second term dominates and $T_\mathrm{c}^{\rm CE1}\sim f^2$. The actual strong disorder scaling $f^{3/2}$ lies between these two. The initial $T_\mathrm{c} \sim f$ scaling at weak disorder, when the 1st order cluster expansion works best, explains why the QMC curves in Fig. 2a of the main text initially rise with decreasing $f$, before the $T_\mathrm{c} \sim f^{3/2}$ scaling takes over.

\subsection{\label{sec:DMFT}Dimer corrections to mean-field}

In this section, rather than capture the leading-order contributions of few-body correlations at all distances as in the cluster expansion, we instead prioritize the pairs of spins that are most strongly correlated. Introducing the quantum correlations of these dimers directly into the mean-field ansatz $H_0$ modifies the $T_\mathrm{c}$ estimate to successfully capture the $f\to0$ dependence on $f$ and $\Delta$.

Accordingly, we work with the modified mean-field Hamiltonian,
\begin{equation}\label{eq:H_d}
    H_d=-h\sum_i s_i^x-\sum_{(ij)\in D}\sum_{\alpha\in\{x,y,z\}}s_i^\alpha J_{ij}^\alpha s_j^\alpha,
\end{equation}
with $J^\alpha_{ij}$ as in $H$~\eqref{eq:H_XXZ} and where we have defined $D$ as the set of dimer pairs, i.e.\ the unordered pairs of spins $(ij)$ which form a dimer. We will return shortly to the exact conditions for assigning these pairs used numerically, but the condition should capture the idea that $(ij)\in D$ if $J_{ij} \gg J_{\rm typ}$. Note that analogous to $H_0$ earlier, for an operator $O$, $\langle O \rangle_d=\tr{\rho_d O}$ and $\langle O \rangle_{d0}=\langle O \rangle_d|_{h=0}$.

The calculation then follows the same as the original mean-field calculation, but with $\chi_{ij}$ no longer purely diagonal due to the dimer correlations. It now has the form,
\begin{equation}\label{eq:chi_ij_dimers}
    \chi_{ij}^{(d)}= \int_0^\beta d\tau \, \langle s_i^x(0)s_j^x(\tau)\rangle_{d0} =\begin{cases}
        \dfrac{1}{J_{ij}(1-\Delta^2)}\dfrac{\Delta(1-x_{ij})+(1+x_{ij}-y_{ij})}{1+x_{ij}+y_{ij}}&(ij)\in D\\
        \dfrac{1}{J_{ik}(1-\Delta^2)}\dfrac{\Delta(1+x_{ik}-y_{ik})+(1-x_{ik})}{1+x_{ik}+y_{ik}}&i=j,\ \exists \, k\text{ such that } (ik)\in D\\
        \dfrac{1}{4}\beta \delta_{ij}&\text{otherwise}
    \end{cases}
\end{equation}
where $x_{ij}=e^{-\beta J_{ij}}$ and $y_{ij}=2e^{-\beta J_{ij}(1-\Delta)/2}$. Importantly, Eq.~\eqref{eq:Tc_MF_condition} must be adapted to account for the exclusion of dimer correlations in $H_{\rm diff}$:
\begin{equation}\label{eq:F_dimers}
    \left.\frac{\partial^2F_d}{\partial h^2}\right|_{h=0}= - \sum_{\substack{ijkl\\(jk)\notin D}} \chi^{(d)}_{ij}\, J_{jk}\, \chi^{(d)}_{kl} + \sum_{ij} \chi^{(d)}_{ij} =0.
\end{equation}

We solve this equation numerically, averaging over disorder realizations the same way as for QMC, with a fixed $N=1000$ spins for all values of $f$. Interactions are again calculated with an Ewald summation. To assign the dimer pairs $D$, we actually pair up all spins, using the same matching algorithm as for the cDTWA clusters. For spins with $J_{ij} \lesssim J_{\rm typ}$, including these terms in $H_d$ makes little difference, so the dominant $\Delta$-dependence still comes from the strongest-coupled dimers.

We also performed a 1st order cluster expansion on top of $H_d$, but after taking $\lambda\to1$, we found that the corrections were not well controlled to 1st order, resulting in poorly behaved $T_\mathrm{c}$ curves.

\subsection{\label{sec:CE2}2nd order cluster expansion}

Returning to the cluster expansion (without the dimer terms), we also show the result of going to 2nd order, which turns out to accurately describe $\Delta_{\rm peak}$ in the weak disorder limit over a range of $f$. This is in contrast to the 1st order calculation, for which $\Delta_{\rm peak}=-0.5$ independent of $f$.

We now must include in addition to Eqs.~\eqref{eq:K_ds} and \eqref{eq:B} the following higher order terms:
\begin{align}
    \frac{1}{\beta^2 N}K_3''|_{h=0}&=\beta^3B_3+6\beta^2B_2+6\beta^3B_1,\label{eq:K3_d}\\
    B_3&=N^{-1}\beta^{-5}\left(A_{3,2}^{(0)}-A_{3,0}^{(0)}A_{0,2}^{(0)}-3A_{3,0}^{(0)}A_{1,2}^{(0)}\right) \label{eq:B3} \\
    &=\frac{1}{256}\left(-6f^3\frac{1}{L^2}\sum_{ijkl\in\Lambda}J_{ij}J_{jk}J_{kl}+2(4+\Delta^3)f^2\frac{1}{L^2}\sum_{ijk\in\Lambda}J_{ij}J_{jk}J_{ki}\right.\notag\\
    &\qquad\qquad\, \left.+4(4+\Delta+\Delta^2)f^2\frac{1}{L^2}\sum_{ijk\in\Lambda}J_{ij}^2J_{jk}-2(3+4\Delta+\Delta^2)\frac{1}{L^2}\sum_{ij\in\Lambda}J_{ij}^3\right).\notag
\end{align}

The last line is derived from summing the contributions from the diagrams shown in Tables \ref{tab:HHHXX} and \ref{tab:HHXHX}, which contribute with equal weight to $\tr{\overline{HHHS^xS^x}}=\frac{1}{2}\tr{HHHS^xS^x}+\frac{1}{2}\tr{HHS^xHS^x}$. The free energy per site is the same as Eq.~\eqref{eq:f_1_d2} but with the 2nd order correction:
\begin{equation}\label{eq:f_2_d2}
    \left.\frac{\partial^2 f_\lambda^{(2)}}{\partial h^2}\right|_{h=0}=\beta/4+\beta^2B_1-\lambda(\beta/2+4\beta^2 B_1 + \beta^3 B_2) + \frac{\lambda^2}{2}(\beta/2+10\beta^2 B_1 + 7\beta^3 B_2 + \beta^4 B_3).
\end{equation}

Solving for the inverse critical temperature, we now find,
\begin{equation}\label{eq:bc_2nd_order}
    \beta_\mathrm{c}^{(\lambda)} = -\frac{1}{4B_1}+\lambda\frac{-8 B_1^2+B_2}{16B_1^3}+ \lambda^2\frac{-128 B_1^4+36B_1^2B_2 - 4B_2^2+B_1B_3}{128 B_1^5} +O(\lambda^3).
\end{equation}
Truncating to $O(\lambda^2)$ and taking $\lambda\to1$, we find this approximation is not well controlled since we do not observe a hierarchy of scales in the coefficients for all $\Delta$. If instead we solve directly for the extremum of Eq.~\eqref{eq:bc_2nd_order} to $O(\lambda^2)$ and then take $\lambda\to1$, the approximation does successfully capture the weak disorder behavior of $\Delta_{\rm peak}$. Doing so, we find:
\begin{equation}\label{eq:D_peak}
    \Delta_{\rm peak}^{\rm CE2}=-\frac{1}{2}+\frac{9}{16}\frac{f^{-1}\frac{1}{L^2}\sum_\Lambda J_{ij}^3 - \frac{1}{L^2}\sum_\Lambda J_{ij}J_{jk}J_{ki}}{\left(\frac{1}{L^2}\sum_\Lambda J_{ij}\right) \left(\frac{1}{L^2}\sum_\Lambda J_{ij}^2\right)}
\end{equation}

\begin{figure}[h]
    \centering
    \includegraphics{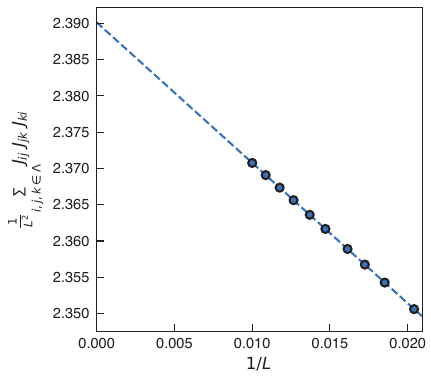}
    \caption{Calculation of $\frac{1}{L^2}\sum_\Lambda J_{ij} J_{jk} J_{ki}$ as $L\to\infty$, which appears in Eq.~\eqref{eq:D_peak}. We use linear extrapolation versus $1/L$ to estimate a value $\approx 2.390$ at the intercept (i.e.\ the thermodynamic limit).}
    \label{fig:Jsum}
\end{figure}

We have already calculated the two sums in the denominator (Eqs.~\eqref{eq:Jij_sum_numeric} and \eqref{eq:Jij_2_sum}), and we can calculate the first sum in the numerator as,
\begin{equation}\label{eq:Jij_3}
    \frac{1}{L^2}\sum_{ij\in\Lambda} J_{ij}^3=4\sum_{n=1}^\infty\sum_{m=0}^\infty
    \left((n^2+m^2)^{-3/2}\right)^3=4.191\ldots.
\end{equation}
We numerically estimate the second sum in the numerator on finite systems with open boundary conditions, and use linear extrapolation versus $1/L$ to find the $1/L\to0$ limit. As shown in Fig.~\ref{fig:Jsum}, we find
\begin{equation}\label{eq:Jijk}
    \frac{1}{L^2}\sum_{ijk\in\Lambda}J_{ij}J_{jk}J_{ki}\approx 2.390.
\end{equation}

Plugging in these numerical values, in Fig.~\ref{fig:delta_peak} we compare $\Delta_{\rm peak}$ from the 2nd order cluster expansion of Eq.~\eqref{eq:D_peak} with that of the dimer corrections to mean-field and of the QMC data. The 2nd order cluster expansion works well in the weak disorder limit, whereas the dimer correction works well for the strong-disorder limit.

\begin{figure}[h]
    \centering
    \includegraphics{figures/S_delta_peak.pdf}
    \caption{Comparison of $\Delta_{\rm peak}$ versus $f$ for approximations. The circles show the data from QMC. The red lines show the cluster expansion (``CE'') to 1st order (dotted red) and 2nd order (solid red). The black dashed line shows the mean-field theory with dimer corrections (``dimer MFT''). The 2nd order cluster expansion works well for the weak disorder limit, whereas the dimer corrections work well for the strong disorder limit.}
    \label{fig:delta_peak}
\end{figure}

\section{Analytic treatment of the strong disorder limit}

In this section, we analytically treat the phase boundary for scalable squeezing in the $f\to0$ asymptotic limit (strong disorder). Specifically, we determine the critical value $\Delta_\mathrm{c}$ above which scalable squeezing is possible. We emphasize that $\Delta_\mathrm{c}$ marks a boundary in squeezing behavior rather than a thermodynamic phase transition, and therefore does not correspond to a critical temperature or energy density such as $T_\mathrm{c}$ or $E_\mathrm{c}$.

Our central result is that as $f\to0$, the strongest-coupled dimers control the change in $E_\mathrm{c}$ with respect to $\Delta$ and lead to $1-\Delta_\mathrm{c} \sim \sqrt{f}$. Thus for any finite $f$, scalable squeezing is possible sufficiently close to the Heisenberg point. Furthermore, in the pre-asymptotic regime, when one accounts for the field on the dimers from the other spins, this phase boundary becomes linear, with $1-\Delta_\mathrm{c} \sim f$, matching the QMC numerics [Fig.~1c of the main text].

\subsection{Asymptotic \texorpdfstring{$\Delta_\mathrm{c} \sim 1-\sqrt{f}$}{Delta\_c \textasciitilde\ 1 - sqrt(f)} behavior}

Our main approximation is a single bond-decimation step: all pairs of spins with separation $r_{ij} < r_{\rm typ}:=a/\sqrt{\pi f}$ [Eq.~\eqref{eq:rtyp_prop}] are treated as isolated subsystems. They form effective spin-$1$ dimers with strong internal couplings $J_{ij} > J_{\rm typ}$, while all remaining sites are monomers. Interactions between these effective spins (dimers or monomers) are $\lesssim J_{\rm typ}$. The intra-dimer bond lengths are distributed according to Eq.~\eqref{eq:Pnn}. In this picture, computing the excitation energy reduces to averaging over the internal spectrum of the dimers and the dipolar mean field generated by the other effective spins.

Following the arguments of the main text, scalable squeezing is possible when $E_\mathrm{c} > E_x$. Moreover, $E_\mathrm{c}$ varies with the anisotropy $\Delta$, whereas $E_x$ does not, and hence critical anisotropy is given by
\begin{equation}
    E_\mathrm{c}(\Delta_\mathrm{c}) = E_x.
\end{equation}
$E_x$ can be estimated in a continuous space approximation by,
\begin{equation}\label{eq:Ex_fto0}
    E_x = -\frac{1}{8}\int_a^\infty 2\pi r \rho \frac{J}{r^3}dr=-\frac{\pi J}{4 a^3}f,
\end{equation}
where as in Sec.~\ref{sec:Ji}, $\rho=f/a^2$ is the spin density and $a$ the lattice constant.

\begin{figure}[h]
    \centering
    \includegraphics{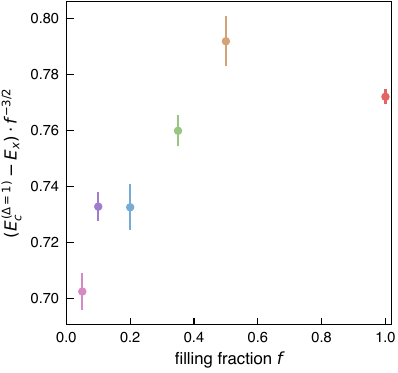}
    \caption{Scaling of $E_\mathrm{c} - E_x$ versus $f$ at $\Delta = 0.9$, near the Heisenberg point. The fact that $(E_\mathrm{c} - E_x)\cdot f^{-3/2}$ approaches a constant value as $f\to0$ indicates that Eq.~\eqref{eq:Heisenberg_excitation_energy} is the correct scaling.}
    \label{fig:Ec_Heisenberg}
\end{figure}

At exactly the Heisenberg point $\Delta = 1$, the $x$-state is actually both the dimer ground state and the many-body ground state. Since at this point the spin-$1$ dimers do not have an internal gap between states (recall $E_{\rm gap} = J_{ij}(1-\Delta)/2$), the excitation energy per spin at $T_\mathrm{c}$ is just the average local field $h$ from the other spins,
\begin{equation}\label{eq:Heisenberg_excitation_energy}
    E_\mathrm{c}^{(\Delta = 1)} - E_x \sim h= \int_{r_{\rm typ}}^\infty 2\pi r \rho \frac{J}{r^3}dr=\frac{2\pi \rho J}{r_{\rm typ}}=2J_{\rm typ} \sim f^{3/2},
\end{equation}
where the lower cutoff at $r_{\rm typ}$ reflects that all shorter-range interactions have already been merged into spin-$1$ dimers. This scaling is directly confirmed by the QMC in Fig.~\ref{fig:Ec_Heisenberg}.

Now consider perturbing $\Delta$ below the Heisenberg point. $E_x$ is unaffected, however the calculation of $E_\mathrm{c}$ is altered. Specifically, at the critical energy density of magnetic ordering, the most strongly coupled dimers will have excitation gaps greater than $T_\mathrm{c}$ and hence be frozen into their ground states 
\begin{equation}\label{eq:Ed}
    E_\mathrm{gs} = -\frac{1}{4}(2-\Delta)J_{ij}.
\end{equation}
The leading $\Delta$-dependence of $E_\mathrm{c}$ arises entirely from the internal energy shift of the dimers: interactions between effective sites remain $\sim J_{\rm typ}$ and therefore still only contribute at $O(f^{3/2})$.

We estimate the change to $E_\mathrm{c}$ as follows. We assume that dimers are frozen at $E_\mathrm{c}$ if they satisfy $E_\mathrm{gap}>T_\mathrm{c}$, which translates to a condition on the dimer separation 
\begin{equation}\label{eq:rF}
    r_{ij} < r_\mathrm{F} := \left(\frac{(1-\Delta) J}{2 T_\mathrm{c}}\right)^{1/3}.
\end{equation}
Summing up the contributions from all such dimers, we obtain the change in $E_\mathrm{c}$:
\begin{equation}
\label{eq:Ec_change}
    \begin{aligned}
    E_\mathrm{c} - E_\mathrm{c}^{(\Delta = 1)} = \frac{1}{2}\int_a^{r_F}2\pi r \rho \left(E_\mathrm{gs}(r) - E_\mathrm{gs}^{(\Delta = 1)}(r)\right)dr=-\frac{\pi J (1-\Delta)f}{4a^3} + O(f^{3/2}),
    \end{aligned}
\end{equation}
where we have assumed $a \ll r_\mathrm{F} \ll r_{\rm typ}$.

Requiring this $O(f)$ shift to cancel the Heisenberg-point excitation energy $E_\mathrm{c} - E_x \sim h\sim f^{3/2}$ fixes the scaling of $\Delta_\mathrm{c}$,
\begin{equation}\label{eq:Delta_c}
    \frac{\pi J (1-\Delta)f}{4a^3} \sim h\implies 1-\Delta_\mathrm{c} \sim \sqrt{f}.
\end{equation}
This is the scaling of the phase boundary in the $f\to0$ limit as illustrated in Fig.~1c of the main text, with a prefactor chosen to match the QMC data point at the lowest $f$.

Note that at $\Delta_\mathrm{c}$ and as $f\to0$, Eq.~\eqref{eq:rF} gives $r_\mathrm{F} \sim af^{-1/3}$, which satisfies $a \ll r_\mathrm{F} \ll r_{\rm typ} \sim a f^{-1/2}$. Thus we confirm the approximations used in evaluating Eq.~\eqref{eq:Ec_change} are consistent.

\subsection{Pre-asymptotic \texorpdfstring{$\Delta_\mathrm{c} \sim 1-f$}{Delta\_c \textasciitilde\ 1 - f} behavior}

We note that a marginally more careful treatment is able to capture the pre-asymptotic $\Delta_\mathrm{c} \sim 1-f$ behavior observed in the numerics [Fig.~1c of the main text]. 

In the pre-asymptotic regime we may model the interactions with other spins as leading to a field $h$ on the dimer, which is not yet dominated by the coupling scale: that is $J_{ij}(1-\Delta)$ and $h$ may be similar scales. Accounting for both scales, the dimer ground state energy is given by
\begin{equation}\label{eq:Ed_field}
    E_\mathrm{gs} = -\left(J_{ij}/4+\sqrt{h^2 + [J_{ij}(1-\Delta)/4]^2}\right).
\end{equation}
We note that the energy change $E_\mathrm{gs} - E_\mathrm{gs}^{(\Delta=1)}$ goes rapidly to zero for $r_{ij} \gg r_{\mathrm{F}}$:
\begin{equation}
    E_\mathrm{gs} - E_\mathrm{gs}^{(\Delta=1)} = O(r_\mathrm{F}^6/r_{ij}^6),
\end{equation}
and consequently it is not necessary to account for freezing by hand, and the energy change at criticality is given by
\begin{equation}
    E_\mathrm{c} - E_\mathrm{c}^{(\Delta=1)} = \frac{1}{2}\int_a^{\infty}2\pi r \rho \left(E_\mathrm{gs}(r) - E_\mathrm{gs}^{(\Delta = 1)}(r)\right)dr = - \frac{1}{2}\pi f h \left( 1 - {}_2F_1\left( -\frac{1}{2},-\frac{1}{3},\frac{2}{3},-\frac{(1-\Delta)^2 J^2}{16 a^6 h^2} \right)   \right)
\end{equation}
where ${}_2F_1$ is the ordinary hypergeometric function.

To leading order at small $h$ we recover the previous scaling~\eqref{eq:Ec_change},
\begin{equation}
    E_\mathrm{c} - E_\mathrm{c}^{(\Delta=1)} \sim -\frac{J (1-\Delta) f}{a^3}.
\end{equation}
However, expanding at small $1-\Delta$ we obtain 
\begin{equation}
    E_\mathrm{c} - E_\mathrm{c}^{(\Delta=1)} \sim -\frac{J^2 (1 - \Delta)^2 f}{h a^6}
\end{equation}
Between these limits, as $1-\Delta$ is increased, $E_\mathrm{c} - E_\mathrm{c}^{(\Delta=1)}$ simply smoothly crosses over from the pre-asymptotic $\sim (1-\Delta)^2$ behavior to the asymptotic $(1-\Delta)$ behavior. 

As before, for squeezing we require that this change to $E_\mathrm{c}$ is not larger than its $\Delta=1$ excitation from $E_x$. Specifically, we are in the squeezing regime $E_x < E_\mathrm{c}$ if
\begin{equation}
     |E_\mathrm{c} - E_\mathrm{c}^{(\Delta=1)}| \lesssim E_\mathrm{c}^{(\Delta=1)} - E_x \sim h
\end{equation}
Setting these terms equal then yields the critical $\Delta_\mathrm{c}$
\begin{equation}
    \frac{J^2 (1 - \Delta_\mathrm{c})^2 f}{h a^6} \sim h \implies 1-\Delta_\mathrm{c} \sim f
\end{equation}
where we have used that at criticality $h \sim J_\mathrm{typ} \sim J f^{3/2}/a^3$.

\begin{table}[h]
\centering
\begin{tabular}{c | c}
    $\begin{diagram}{8}
        \encircle{2}{3}
        \encircle{4}{5}
        \encircle{6}{7}
        \disconnect{1}{8}
        \disconnect{1}{6}
        \disconnect{8}{3}
        \connect{1}{2}
        \connect{3}{4}
        \connect{5}{6}
        \connect{7}{8}
    \end{diagram} \times 48\ $ &
    \(\displaystyle 
    -\frac{1}{256}6\left(\sum_{ijkl}J_{ij}J_{jk}J_{kl}-2\sum_{ijk}J_{ij}^2J_{jk} + \sum_{ij}J_{ij}^3 - \sum_{ijk}J_{ij}J_{jk}J_{ki}\right)
    \)
    \\
    [5ex]\hline\rule{0pt}{9ex}
    $\begin{diagram}{8}
        \encircle{2}{3}
        \encircle{4}{5}
        \encircle{6}{7}
        \disconnect{1}{8}
        \connect{7}{8}
        \connect{1}{2}
        \connect{2}{4}
        \connect{4}{1}
        \connect{3}{5}
        \connect{5}{6}
        \connect{6}{3}
    \end{diagram} \times 4\ $ &
    \(\displaystyle 
    \frac{1}{256}4\Delta\left(\sum_{ijk}J_{ij}^2J_{jk} - \sum_{ij}J_{ij}^3\right)
    \)
    \\
    [5ex]\hline\rule{0pt}{9ex}
    $\begin{diagram}{8}
        \encircle{2}{3}
        \encircle{4}{5}
        \encircle{6}{7}
        \connect{2}{7}
        \connect{1}{3}
        \connect{3}{4}
        \connect{4}{1}
        \connect{8}{5}
        \connect{5}{6}
        \connect{6}{8}
    \end{diagram} \times 16\ $ &
    \(\displaystyle 
    \frac{1}{256}2\Delta(1+\Delta)\sum_{ijk}J_{ij}J_{jk}J_{ki}
    \)
    \\
    [5ex]\hline\rule{0pt}{9ex}
    $\begin{diagram}{8}
        \encircle{2}{3}
        \encircle{4}{5}
        \encircle{6}{7}
        \connect[2]{8}{5}
        \connect[2]{7}{2}
        \connect{1}{4}
        \connect{3}{6}
        \disconnect{3}{4}
    \end{diagram} \times 8\ $ &
    \(\displaystyle 
    \frac{1}{256}4(1+\Delta^2)\left(\sum_{ijk}J_{ij}^2J_{jk}-\sum_{ij}J_{ij}^3\right)
    \)
    \\
    [5ex]\hline\rule{0pt}{9ex}
    $\begin{diagram}{8}
        \encircle{2}{3}
        \encircle{4}{5}
        \encircle{6}{7}
        \connect[2]{8}{5}
        \connect[2]{7}{2}
        \connect[3]{1}{4}
        \connect[3]{3}{6}
        \disconnect{3}{4}
    \end{diagram} \times -2\ $ &
    \(\displaystyle 
    \frac{1}{256}2\sum_{ij}J_{ij}^3
    \)
\end{tabular}
\caption{Diagrams for $\tr{HHHS^xS^x}$ that contribute to $A^{(0)}_{3,2}$ [Eq.~\eqref{eq:A_nm_0}], with equivalent diagrams shown by one representative on the left, and the total from all equivalent diagrams on the right. Note that generally the sum over spin components is included in the diagram's evaluation in the right column, but not the counting in the left column. There are exceptions where permuting the spin components between $H$'s (ellipses) results in exact cancellations, which is included in the counting on the left. In any case, the sum over the rightmost column gives the total value. The last row is needed because of the overcounting of this diagram when $3A_{3,0}^{(0)}A_{1,2}^{(0)}$ is subtracted, since the diagram is counted three times by that term but once by $A_{3,2}^{(0)}$.}\label{tab:HHHXX}
\end{table}

\begin{table}[h]
\centering
\begin{tabular}{c | c}
    $\begin{diagram}{8}
        \encircle{1}{2}
        \encircle{4}{5}
        \encircle{7}{8}
        \disconnect{3}{6}
        \disconnect{4}{6}
        \disconnect{3}{5}
        \connect{1}{4}
        \connect{2}{3}
        \connect{5}{8}
        \connect{6}{7}
    \end{diagram} \times 48\ $ &
    \(\displaystyle 
    -\frac{1}{256}6\left(\sum_{ijkl}J_{ij}J_{jk}J_{kl}-2\sum_{ijk}J_{ij}^2J_{jk} + \sum_{ij}J_{ij}^3 - \sum_{ijk}J_{ij}J_{jk}J_{ki}\right)
    \)
    \\
    [5ex]\hline\rule{0pt}{9ex}
    $\begin{diagram}{8}
        \encircle{1}{2}
        \encircle{4}{5}
        \encircle{7}{8}
        \connect{1}{8}
        \connect{8}{5}
        \connect{5}{1}
        \connect{2}{3}
        \connect{3}{4}
        \connect{4}{2}
        \connect{6}{7}
        \disconnect{3}{7}
    \end{diagram} \times 4\ $ &
    \(\displaystyle 
    \frac{1}{256}4\Delta\left(\sum_{ijk}J_{ij}^2J_{jk} - \sum_{ij}J_{ij}^3\right)
    \)
    \\
    [5ex]\hline\rule{0pt}{9ex}
    $\begin{diagram}{8}
        \encircle{1}{2}
        \encircle{4}{5}
        \encircle{7}{8}
        \connect{1}{8}
        \connect{2}{3}
        \connect{3}{4}
        \connect{4}{2}
        \connect{5}{6}
        \connect{6}{7}
        \connect{7}{5}
    \end{diagram} \times 16\ $ &
    \(\displaystyle 
    -\frac{1}{256}2\Delta(1+\Delta)\sum_{ijk}J_{ij}J_{jk}J_{ki}
    \)
    \\
    [5ex]\hline\rule{0pt}{9ex}
    $\begin{diagram}{8}
        \encircle{1}{2}
        \encircle{4}{5}
        \encircle{7}{8}
        \connect{1}{8}
        \connect{2}{4}
        \connect[2]{5}{7}
        \connect[2]{3}{6}
    \end{diagram} \times 16\ $ &
    \(\displaystyle 
    \frac{1}{256}4(1+\Delta^3)\sum_{ijk}J_{ij}J_{jk}J_{ki}
    \)
    \\
    [5ex]\hline\rule{0pt}{9ex}
    $\begin{diagram}{8}
        \encircle{1}{2}
        \encircle{4}{5}
        \encircle{7}{8}
        \connect{1}{5}
        \connect{5}{7}
        \connect{7}{1}
        \connect[2]{2}{4}
        \connect[2]{4}{8}
        \connect[2]{8}{2}
        \connect[2]{3}{6}
    \end{diagram} \times 16\ $ &
    \(\displaystyle 
    -\frac{1}{256}8\Delta\sum_{ij}J_{ij}^3
    \)
    \\
    [5ex]\hline\rule{0pt}{9ex}
    $\begin{diagram}{8}
        \encircle{1}{2}
        \encircle{4}{5}
        \encircle{7}{8}
        \connect[2]{1}{6}
        \connect[2]{4}{8}
        \connect{2}{3}
        \connect{5}{7}
    \end{diagram} \times 16\ $ &
    \(\displaystyle 
    \frac{1}{256}4(1+\Delta^2)\sum_{ijk}J_{ij}^2J_{jk}
    \)
    \\
    [5ex]\hline\rule{0pt}{9ex}
    $\begin{diagram}{8}
        \encircle{2}{3}
        \encircle{4}{5}
        \encircle{6}{7}
        \connect[2]{8}{5}
        \connect[2]{7}{2}
        \connect[3]{1}{4}
        \connect[3]{3}{6}
        \disconnect{3}{4}
    \end{diagram} \times -2\ $ &
    \(\displaystyle 
    \frac{1}{256}2\sum_{ij}J_{ij}^3
    \)
\end{tabular}
\caption{Diagrams for $\tr{HHS^xHS^x}$, shown the same way as in Table~\ref{tab:HHHXX}.}\label{tab:HHXHX}
\end{table}

\FloatBarrier
\newpage

\bibliography{refs}